\documentclass[11pt, a4paper]{article}
\pdfoutput=1
\usepackage{jcappub}

\usepackage[usenames, dvipsnames]{xcolor}
\usepackage{bbold}
\usepackage{booktabs}
\usepackage{multirow}
\usepackage{amssymb}
\usepackage{amsmath}
\usepackage[export]{adjustbox}
\usepackage{floatrow}
\newfloatcommand{capbtabbox}{table}[][3.5in]
\newfloatcommand{capbfigbox}{figure}[][2.3in]
\usepackage{blindtext}
\usepackage{booktabs}
\usepackage{aas_macros}
\usepackage{slashed}

\newcommand{\be}{\begin{equation}}
\newcommand{\ee}{\end{equation}}
\newcommand{\nl}{\nonumber \\}
\newcommand{\x}{\chi}
\newcommand{\xii}{\chi_i}
\newcommand{\xiibar}{\overline{\chi}_i}
\newcommand{\xjj}{\chi_j}
\newcommand{\xjjbar}{\overline{\chi}_j}
\newcommand{\xI}{\chi_1}
\newcommand{\xII}{\chi_2}
\newcommand{\xIII}{\chi_3}
\newcommand{\xp}{\chi^+}
\newcommand{\xpbar}{\overline{\chi^+}}
\newcommand{\xm}{\chi^-}

\newcommand{\xpp}{\chi^{++}}
\newcommand{\xppbar}{\overline{\chi^{++}}}

\usepackage[usenames,dvipsnames]{xcolor}

\def\lsim{\mathrel{\raise.3ex\hbox{$<$\kern-.75em\lower1ex\hbox{$\sim$}}}}
\def\gsim{\mathrel{\raise.3ex\hbox{$>$\kern-.75em\lower1ex\hbox{$\sim$}}}}

\definecolor{paleblue}{rgb}{0.69, 0.93, 0.93}  

\begin{document}

\hspace*{110mm}{\large \tt FERMILAB-PUB-16-111-A-T}

\vskip 0.2in

\title{Mixed Dark Matter in Left-Right Symmetric Models}

\author[a]{Asher Berlin,}
\emailAdd{berlin@uchicago.edu}

\author[b]{Patrick J. Fox,}
\emailAdd{pjfox@fnal.gov}

\author[c,d]{Dan Hooper,}
\emailAdd{dhooper@fnal.gov}

\author[c,e]{and Gopolang Mohlabeng}
\emailAdd{gopolang.mohlabeng@ku.edu}

\affiliation[a]{Department of Physics, Enrico Fermi Institute, University of Chicago, Chicago, IL}
\affiliation[b]{Fermi National Accelerator Laboratory, Theory Group, Batavia, IL}
\affiliation[c]{Fermi National Accelerator Laboratory, Center for Particle
Astrophysics, Batavia, IL}
\affiliation[d]{University of Chicago, Department of Astronomy and Astrophysics, Chicago, IL}
\affiliation[e]{University of Kansas, Department of Physics and Astronomy, Lawrence, KS}

\abstract{Motivated by the recently reported diboson and dijet excesses in Run 1 data at ATLAS and CMS, we explore models of mixed dark matter in left-right symmetric theories. In this study, we calculate the relic abundance and the elastic scattering cross section with nuclei for a number of dark matter candidates that appear within the fermionic multiplets of left-right symmetric models. In contrast to the case of pure multiplets, WIMP-nucleon scattering proceeds at tree-level, and hence the projected reach of future direct detection experiments such as LUX-ZEPLIN and XENON1T will cover large regions of parameter space for TeV-scale thermal dark matter. Decays of the heavy charged $W^\prime$ boson to particles in the dark sector can potentially shift the right-handed gauge coupling to larger values when fixed to the rate of the Run 1 excesses, moving towards the theoretically attractive scenario, $g_R = g_L$. This region of parameter space may be probed by future collider searches for new Higgs bosons or electroweak fermions.}

\maketitle

\section{Introduction}

The ATLAS collaboration has recently reported an excess of events consistent with those arising from an approximately 2 TeV resonance decaying to a pair of Standard Model (SM) gauge bosons.  If interpreted as a $WZ$ final state, this excess has a local significance of 3.4$\sigma$, or 2.5$\sigma$ after taking into account an appropriate trials factor~\cite{Aad:2015owa}. The possibility that this excess could be associated with new physics is strengthened by the results of Run 1 dijet searches at CMS~\cite{Khachatryan:2015sja} and ATLAS~\cite{Aad:2014aqa}, each of which report a modest excess (2.2 and 1.0$\sigma$, respectively) at a similar mass of approximately 1.8 TeV.  In addition, the CMS collaboration reports 2.1 and 1.5$\sigma$ excesses in their searches for leptonically-tagged resonances decaying to $HW$~\cite{CMS:2015gla} or to gauge bosons~\cite{Khachatryan:2014gha}, respectively, both at approximately 1.8 TeV. 

These anomalies have renewed interest in models with a new charged gauge boson, $W'$, with a mass of approximately 1.8 to 2 TeV, including those predicted within the context of left-right symmetric models~\cite{Dobrescu:2015qna,Brehmer:2015cia,Dev:2015pga,Heikinheimo:2014tba,Das:2016akd}. Such scenarios have long been perceived as theoretically attractive~\cite{Pati:1974yy,Mohapatra:1974hk,Mohapatra:1974gc,Senjanovic:1975rk,Marshak:1979fm,Mohapatra:1980qe,Mohapatra:1986uf}, and can emerge naturally within the context of Grand Unified Theories based on $SO(10)$ or $E_6$~\cite{Langacker:1980js,Hewett:1988xc,Boucenna:2015sdg}. At low energies, left-right symmetric models are described by the gauge group $SU(3)_c \times SU(2)_L\times SU(2)_R \times U(1)_{B-L}$, and thus include new massive gauge bosons, $W'$ and $Z'$, that couple to right-handed fermions, including right-handed neutrinos. To be phenomenologically viable, left-right symmetric models also require an extended Higgs sector, and often include additional fermionic particle content. It has been suggested that the modest excess of same-sign dilepton events with multiple $b$-jets observed at ATLAS~\cite{Aad:2015gdg} could be explained by the Higgs sector of such a model, while a left-right symmetric model with TeV-scale right-handed neutrinos could account for the CMS excess of $e^+e^-$ events that include a pair of jets with an invariant mass of $\sim$2 TeV~\cite{Coloma:2015una}. Although the first 13 TeV data from the LHC have not been definitive~\cite{CMS-PAS-EXO-15-002,ATLAS-CONF-2015-068,ATLAS-CONF-2015-071,ATLAS-CONF-2015-075}, the ongoing run at the LHC is expected to reach the sensitivity required to conclusively test these models over the coming year~\cite{Dobrescu:2015qna}.

There are a number of potentially viable dark matter candidates that one can identify within the context of left-right symmetric models~\cite{Heeck:2015qra,Brehmer:2015cia,Garcia-Cely:2015quu,Nemevsek:2012cd,Bezrukov:2009th}, and their supersymmetric extensions~\cite{Esteves:2011gk,An:2011uq,Bhattacharya:2013nya}. In this paper, we limit our scope to non-supersymmetric models, considering a wide range of dark matter candidates contained within various $SU(2)_{L,R}$ fermion multiplets and their mixtures. In some respects, this follows the previous work of Heeck and Patra, who considered dark matter candidates 
that are members of left-right fermion triplets or quintuplets~\cite{Heeck:2015qra}. More recently, Garcia-Cely and Heeck extended this approach by considering those candidates found within fermion bidoublets or bitriplets, or scalar doublets or 7-plets~\cite{Garcia-Cely:2015quu}. In this study, we build upon this earlier work by considering dark matter candidates found within the fermion multiplets of a left-right symmetric model, but without restricting ourselves to pure states. In particular, we find that fermion singlet-triplet, singlet-bidoublet, and triplet-bidoublet mixtures each constitute phenomenologically viable dark matter candidates. Furthermore, we show that such states are automatically cosmologically stable, without the need for any additional parity or symmetry. In contrast to pure states, mixed dark matter in left-right symmetric models can undergo significant scattering with nuclei, potentially falling within the reach of direct detection experiments such as LUX, LUX-ZEPLIN (LZ), and XENON1T. Additionally, whereas the mass splitting between the neutral and charged particles of the dark sector is fixed when considering pure states, this splitting can be adjusted more freely in mixed models, allowing us, for example, to turn on or off the effects of coannihilation in the early universe. 

Although we focus this study on the parameter space motivated by the diboson excess ($m_{W'}\sim 2$ TeV, $g_R \sim 0.5$), we note that dark matter within the context of left-right symmetric models would remain interesting even in the absence of such a putative signal. With this in mind, we have presented many of our results in a way that can be straightforwardly applied to other scenarios within the larger parameter space of left-right symmetric models. 

The remainder of this paper is structured as follows. In Secs.~\ref{higgs} and~\ref{gauge}, we discuss the Higgs and gauge sectors of left-right symmetric models, respectively, describing their particle content and interactions. In Sec.~\ref{sec:mixedmodels}, we consider three scenarios in which the dark matter is a mixture of fermions found in $SU(2)$ singlets, bidoublets and triplets, in each case finding regions of parameter space that predict an acceptable thermal relic abundance and that are consistent with the constraints from direct detection experiments. In Sec.~\ref{summary}, we summarize our results and conclusions. In a series of Appendices, we provide expressions for many of the interactions predicted within this class of models, along with general results for loop induced mass corrections to neutral and charges states in the dark sector.

\section{The Higgs Sector}
\label{higgs}

The spontaneous symmetry breaking of $SU(2)_L \times SU(2)_R \times U(1)_{B-L}$ down to $U(1)_{EM}$ requires an extended Higgs sector (for a review, see Ref.~\cite{Duka:1999uc}). In particular, the minimal content of a left-right symmetric model includes a complex scalar triplet with quantum numbers $\Delta_R : ({\bf 1},{\bf 3},2)$, and a complex scalar bidoublet with quantum numbers $\phi : ({\bf 2},{\bf 2},0)$. 

The electric charge of a given state is defined in relation to its weak isospin and $B-L$ quantum numbers: 
\be
Q = T_{3L} + T_{3R} + \frac{{B-L}}{2} \, .
\ee 
This can be generalized further for triplets and bidoublets, respectively, according to the following:
\begin{eqnarray}
\label{eq:charge}
Q {\bf T} &=& \Big[ \frac{1}{2} \sigma_3 , {\bf T} \Big] + \frac{{B-L}}{2} \, {\bf T}, \nonumber \\
Q {\bf B} &=& \Big[ \frac{1}{2} \sigma_3 , {\bf B} \Big] + \frac{{B-L}}{2} \, {\bf B},
\end{eqnarray}
where ${\bf T}$ and ${\bf B}$ are $2 \times 2$ matrices in which a generic triplet or bidoublet can be embedded. The matrix ${\bf T}$ is further constrained to be traceless.

The charge conjugates of a triplet and a bidoublet, which we will use in Sec.~\ref{sec:mixedmodels}, are defined as:
\begin{eqnarray}
\tilde{{\bf T}} &\equiv& \sigma^2 {\bf T}^* \sigma^2  = - {\bf T}^\dagger, \nonumber  \\
\tilde{{\bf B}} &\equiv& \sigma^2 {\bf B}^* \sigma^2.
\end{eqnarray}

The right-handed Higgs triplet, $\Delta_R$, breaks $SU(2)_L \times SU(2)_R \times U(1)_{B-L}$ down to $SU(2)_L \times U(1)_Y$ after acquiring a vacuum expectation value (VEV), $v_R$. Subsequently, the Higgs bidoublet, $\phi$, breaks $SU(2)_L \times U(1)_Y$ down to $U(1)_{EM}$. The low-energy Higgs potential of this model corresponds to a restricted form of a two-Higgs doublet model (2HDM)~\cite{Dobrescu:2015yba}. Working in unitary gauge and the alignment limit (in which the lightest Higgs is SM-like), we parametrize these Higgs bosons as follows:
\begin{eqnarray}
\Delta_R &=& \begin{pmatrix}  \frac{g_R}{\sqrt{2}~g_L} ~ \frac{m_W}{m_{W^\prime}} ~ c_{2 \beta} H^+   & \Delta^{++} \\ v_R + \frac{1}{\sqrt{2}}~\Delta^0  &   - \frac{g_R}{\sqrt{2}~g_L} ~ \frac{m_W}{m_{W^\prime}} ~ c_{2 \beta} H^+   \end{pmatrix}, 
 \\ \nonumber 
 \\
\phi &=&  \begin{pmatrix} c_\beta v + \frac{1}{\sqrt{2}} ( c_\beta h + s_\beta H + i s_\beta A )& c_\beta H^+  \\ s_\beta H^-  & s_\beta v + \frac{1}{\sqrt{2}} (s_\beta h - c_\beta H + i c_\beta A) \end{pmatrix}, \nonumber 
\end{eqnarray}
where $h$, $H$, $A$, and $H^{\pm}$ represent the Higgs bosons found within a generic 2HDM, while $\Delta^0$ and $\Delta^{++}$ denote the physical right-handed neutral and doubly charged scalars, respectively. The quantities $g_R$ and $g_L$ are the gauge couplings associated with $SU(2)_R$ and $SU(2)_L$, respectively, while $v = 174$ GeV is the SM Higgs VEV. $c_{N\beta}$, $s_{N\beta}$ are the cosine and sine of $N \times \beta$, where $\tan{\beta}$ is the ratio of the two neutral VEVs of $\phi$, analogous to that of a 2HDM. To match the observed rate and mass of the diboson excess, we require $v_R \sim 3-4$ TeV and $g_R \sim 0.5$ \cite{Dobrescu:2015jvn}.  In addition, matching to the observed $W^\prime \to W Z$ rate also requires $0.5 \lesssim \tan{\beta} \lesssim 2$. The masses of $H, A$, and $H^\pm$ naturally take on a common value that can lie anywhere between the weak scale and $v_R \, $. 

The only renormalizable gauge invariant interactions between SM quarks and the extended Higgs sector are given by:

\be
\label{eq:ferm1}
-\mathcal{L} \supset \bar{Q}_L  \left( y ~ \phi + \tilde{y} ~ \tilde{\phi} \right) Q_R + \text{h.c.}
~,
\ee
which after electroweak symmetry breaking (EWSB), gives rise to the following quark mass terms:
\be
-\mathcal{L} \supset (y ~ c_\beta + \tilde{y} ~ s_\beta) ~v~ \bar{u} u + (y ~ s_\beta + \tilde{y} ~ c_\beta) ~v~ \bar{d} d
~.
\ee
A modest tuning of these parameters is required to explain the hierarchy between the top and bottom quark masses.  For $\tan\beta=0.5$ or 2, the Yukawa couplings must be tuned at approximately the 2\% level (allowing for a cancellation between $y s_\beta$ and $\tilde{y} c_\beta$).

Similar terms can be written for the lepton sector, but with an additional coupling to the triplet Higgs:
\be
- \mathcal{L} \supset y_M ~ \overline{(L_R)^c} ~ i \sigma^2 \Delta_R ~ L_R + \text{h.c.}~.
\ee
After EWSB, this term gives a Majorana mass to the right-handed neutrinos and also introduces an interaction of the form $\Delta^{++} ~ \overline{(l_R)^c} ~ l_R$.

Expanding Eq.~(\ref{eq:ferm1}) in terms of the physical Higgs states, we can write the heavy Higgs interactions with SM fermions as follows:
\begin{align}
\mathcal{L} &\supset \frac{m_d - m_u s_{2 \beta}}{\sqrt{2} ~ v c_{2 \beta}} ~ H ~ \bar{u} u + \frac{m_u - m_d s_{2 \beta}}{\sqrt{2} ~ v c_{2 \beta}} ~ H ~ \bar{d} d - \frac{m_l t_{2 \beta}}{\sqrt{2} ~ v} ~ H ~ \bar{l} l
\nl
&+ \frac{m_d - m_u s_{2 \beta}}{\sqrt{2} ~ v c_{2 \beta}} ~ A ~ \bar{u} \, i \gamma^5 u - \frac{m_u - m_d s_{2 \beta}}{\sqrt{2} ~ v c_{2 \beta}} ~ A ~ \bar{d} \, i \gamma^5 d + \frac{m_l t_{2 \beta}}{\sqrt{2} ~ v} ~ A  ~ \bar{l}\,  i \gamma^5 l
\nl
&+ \bigg\{ \frac{1}{v ~ c_{2 \beta}} ~ H^+ ~ \bar{u} \left[ - \left( m_u - m_d s_{2 \beta}\right) P_R + \left( m_d - m_u s_{2 \beta}\right) P_L \right] d + \frac{m_l t_{2 \beta}}{v} ~ H^+ ~ \bar{\nu} ~ P_R ~ l + \text{h.c.} \bigg\}~.
\label{eq:yukawas}
\end{align}

The renormalizable interaction in Eq.~(\ref{eq:ferm1}) couples both Higgs doublets within $\phi$ to up-type and down-type quarks, and can lead to flavor changing couplings through tree-level exchange of the heavy neutral Higgs bosons.  For heavy Higgses above a few TeV in mass, however, these flavor constraints can be avoided \cite{Blanke:2011ry}.  Alternatively, the Yukawa couplings of the quarks can be made to be those of a Type II 2HDM if the renormalizable couplings are small and instead the quarks acquire a mass through higher dimensional operators involving the triplet Higgs \cite{Dobrescu:2015yba}.  We will consider Higgs couplings as in Eq.~(\ref{eq:ferm1}) and assume that all flavor constraints are satisfied.

\section{The Gauge Boson Sector}
\label{gauge}

The gauge bosons acquire masses from the VEVs of the Higgs triplet, $\Delta_R$, and bidoublet, $\phi \, $. Defining $W_{L,R}^{\pm \mu} \equiv \frac{1}{\sqrt{2}} \left( W_{L,R}^{1 \mu} \mp i W_{L,R}^{2 \mu} \right)$, the mass matrices for these states are given as follows:
\be
\mathcal{L} \supset \begin{pmatrix} W_L^{+ \mu} & W_R^{+ \mu} \end{pmatrix} \begin{pmatrix} ~ \frac{1}{2} ~ g_L^2 ~ v^2  & & - \frac{1}{2} ~ g_L ~  g_R ~ s_{2 \beta} ~ v^2 ~  \\ \\ ~ - \frac{1}{2} ~ g_L g_R ~ s_{2 \beta} ~ v^2   & & g_R^2 ~ \left( v_R^2 + \frac{1}{2} v^2 \right) ~ \end{pmatrix} \begin{pmatrix} W_{L \mu}^- \\ \\ W_{R \mu}^- \end{pmatrix}
,
\ee
and
\be
\mathcal{L} \supset \frac{1}{2} \begin{pmatrix} W_{L \mu}^3 & W_{R \mu}^3 & B_\mu \end{pmatrix} \begin{pmatrix} ~\frac{1}{2}~g_L^2~v^2 & - \frac{1}{2} ~ g_L~g_R~v^2 & 0~ \\ \\ ~ - \frac{1}{2} ~ g_L ~ g_R ~ v^2 & \frac{1}{2} ~ g_R^2 ~ v^2 + 2 ~ g_R^2 ~ v_R^2 ~ & -2 ~ g_{_{B-L}} ~ g_R ~ v_R^2 ~ \\ \\ ~0 & -2 ~ g_{_{B-L}} ~ g_R ~ v_R^2 & 2 ~ g_{_{B-L}}^2 ~ v_R^2~ \end{pmatrix} \begin{pmatrix} W_{L \mu}^3 \\ \\ W_{R \mu}^3 \\ \\ B_\mu \end{pmatrix}
,
\ee
where $g_{_{B-L}}$ and $B_\mu$ are the $U(1)_{B-L}$ gauge coupling and field.

Diagonalizing the $W_{L,R}^\pm$ mass matrix in the $v_R \gg v$ limit yields two charged gauge bosons of mass $m_W = g_L v/\sqrt{2}$ and $m_{W^\prime} = g_R v_R \, $. The mixing matrix between these states is given by:
\be
 \begin{pmatrix} W_{L \mu}^\pm \\ \\  W_{R \mu}^\pm \end{pmatrix} = \begin{pmatrix} \cos{\theta_+} & -\sin{\theta_+} \\ \\ \sin{\theta_+} & \cos{\theta_+} \end{pmatrix} \begin{pmatrix} W_\mu^\pm \\ \\  W_\mu^{\prime \pm} \end{pmatrix}
,
\ee
such that
\be 
\sin{\theta_+} \equiv \frac{g_R}{g_L} \left( \frac{m_W}{m_{W^\prime}} \right)^2 s_{2 \beta} 
~.
\ee
The form of $m_W$ given above implies that $g_L$ should be identified with the SM gauge coupling, $g$. 

Diagonalizing the $W_{L}^3$, $W_{R}^3$, $B$ mass matrix yields three neutral gauge bosons with the following masses (again, in the $v_R \gg v$ limit):
\be
m_A^2 = 0~,~~~~ m_Z^2 = \left( g_L^2 + \frac{g_R^2 g_{_{B-L}}^2}{g_R^2 + g_{_{B-L}}^2} \right) \frac{v^2}{2}~,~~~~ m_{Z^\prime}^2 = 2 \left( g_R^2 + g_{_{B-L}}^2 \right) v_R^2 
.
\ee
Comparing the expression for $m_Z$ to that found in the SM, $m_Z^2 = \left( g_L^2 + g'^{2} \right) \frac{v^2}{2}$, we arrive at the following \emph{definition} for the SM hypercharge gauge coupling:
\be
g^\prime \equiv \frac{g_R ~ g_{_{B-L}}}{\sqrt{g_R^2 +g^2_{_{B-L}}}} \, .
\ee
Consistency with the SM requires $g_L \approx 0.65$ while fixing to the diboson rate requires $g_R \approx 0.45-0.6$. Together, these in turn imply $g_{_{B-L}} \approx 0.45-0.6$. 

The mass eigenstates, keeping leading order terms in $v/v_R \, $, are given by:
\be
\begin{pmatrix} W_{L \mu}^3 \\  W_{R \mu}^3 \\ B_\mu \end{pmatrix} = \begin{pmatrix} s_w & c_w & - \frac{g_R}{2 g_L} c_R^3 \left( \frac{m_W}{m_{W^\prime}}\right)^2 \\  c_w s_R & -s_w s_R & c_R \\ c_w c_R & - s_w c_R & -s_R \end{pmatrix} \begin{pmatrix} A_\mu \\ Z_\mu \\ Z_\mu^\prime \end{pmatrix}
~,
\ee
where we have defined 
\be
s_w \equiv \sin{\theta_w} \equiv \frac{g^\prime}{\sqrt{g_L^2+g^{\prime 2}}} ~,~~~ c_w \equiv \cos{\theta_w} ~,~~~ s_R \equiv \sin{\theta_R} \equiv \frac{g^\prime}{g_R} ~,~~~  c_R \equiv \cos{\theta_R}
~.
\ee

In Appendix~\ref{appA}, we provide expressions for the couplings of the $Z^\prime$ and $W^\prime$ to SM fermions, the cubic self-interaction terms involving non-SM gauge bosons, and the non-SM cubic gauge-Higgs interactions terms.  In what follows, we fix the parameters of the new physics to values that fit the excess in the diboson, and related, channels.  In particular, we take the $SU(2)_R$ gauge coupling $g_R=0.45$ and $M_{W^\prime}=1.9$ TeV, which leads to $M_{Z^\prime}=4.4$ TeV.  We take the ratio of the VEVs in the bidoublet scalar to be $\tan\beta=2$, and we assume that the physical scalars in the Higgs triplet are sufficiently heavy such that they take no part in the dynamics.  Although it has been shown \cite{Dobrescu:2015jvn} that if some of the right-handed neutrinos have mass around 1.4--1.7 TeV then their 3-body decay to $ejj$ can explain a CMS excess in $e^+e^-jj$ final state \cite{Khachatryan:2014dka}, we choose here, for simplicity, to decouple these states.  Keeping them in the mass range necessary to explain the $e^+e^-jj$ excess would increase the $W^\prime$ and $Z^\prime$ widths by less than 10\%, which has a small affect on the dark matter relic abundance calculation.

\section{The Dark Matter Sector}
\label{sec:mixedmodels}

\begin{table}[t]
\centering
\begin{tabular}{| c | c | c |}
\hline
$SU(2)$ Fields &  Mixing Possible? \\ \hline \hline
singlet-doublet  & 
$\times$\\ \hline
singlet-triplet & 
$\checkmark$ \\ \hline
singlet-bidoublet  & 
$\checkmark$ \\ \hline
doublet-triplet &  
$\times$\\ \hline
doublet-bidoublet  & 
$\times$\\ \hline
triplet-bidoublet &  
$\checkmark$ \\ \hline
\end{tabular}
\caption{Whether or not mixing is possible through renormalizable Yukawa couplings to the bidoublet and triplet Higgs bosons. Combinations that include multiplets larger than those listed are not able to mix.}
\label{table:mixing}
\end{table}

Previous studies of dark matter in (non-supersymmetric) left-right symmetric models have focused on dark matter composed of pure multiplets. For example, the authors of Ref.~\cite{Heeck:2015qra} considered dark matter candidates that are members of a left-right fermion triplet or quintuplet, while Ref.~\cite{Garcia-Cely:2015quu} extended this to include those states found within a fermion bidoublet or bitriplet, or a scalar doublet or 7-plet. Such candidates can closely resemble what is sometimes referred to as ``minimal dark matter''~\cite{Cirelli:2005uq,Cirelli:2007xd,Cirelli:2009uv,DelNobile:2015bqo}. In this study, we extend the analysis to a wider range of scenarios by considering models in which the dark matter candidate is not necessarily in a pure state, but may instead be a fermion that is a mixture of two or more multiplets, similar to neutralinos in supersymmetry. For implementation of this class of models in regards to the recently reported 750 GeV diphoton excess, see e.g., Ref.~\cite{Berlin:2016hqw}.

Although we restrict our analysis to fermionic dark matter (motivated, in part, by supersymmetric completions of left-right symmetric models~\cite{Demir:2006ef}), we consider arbitrary combinations of fermion multiplets. Mixing between the fermions is induced through the coupling to a bidoublet or triplet Higgs, $\phi$ and $\Delta_R \, $.  At the renormalizable level, gauge invariance allows only a finite set of possible combinations, which involves only singlets, $SU(2)$ doublets, bidoublets, and $SU(2)_R$ triplets.  There are no combinations that include a higher multiplet.  In Table~\ref{table:mixing} we list all possible combinations of distinct fermion representations that can mix via renormalizable Yukawa couplings to the bidoublet and triplet Higgses.

In light of these considerations, we restrict our analysis to the following three mixed cases: singlet-triplet, singlet-bidoublet, and triplet-bidoublet dark matter. In many ways, these are phenomenologically analogous to bino-wino, bino-Higgsino, and wino-Higgsino dark matter in the MSSM, respectively. In the following three subsections, we will discuss each of these cases in turn.

\subsection{Singlet-Triplet Dark Matter}
\label{sec:singlettriplet}

\begin{table}[t]
\centering
\begin{tabular}{| c || c | c | c |}
\hline
Field & Charges & Spin \\ \hline \hline
$S$ & $({\bf 1},{\bf 1},0)$ & 1/2 \\ \hline
$T_1$ & $({\bf 1}, {\bf 3},2)$ & 1/2 \\ \hline
$T_2$ & $({\bf 1},{\bf 3},-2)$ & 1/2 \\ \hline
\end{tabular}
\caption{The $SU(2)_L$, $SU(2)_R$, and $B-L$ charge assignments in the singlet-triplet model.  All fields are colorless.}
\label{table:singlettriplet}
\end{table}

In this scenario, we introduce three Weyl fermions: a singlet, $S$, and two triplets, $T_{1,2}$, with charge assignments as given in Table~\ref{table:singlettriplet}. The triplets are each given $B-L$ charge so that they can couple to the triplet Higgs, $\Delta_R$. Two triplets are needed for anomaly cancellation, which also allows a bare triplet mass term. Note that the presence of $SU(2)_R$ triplets without corresponding $SU(2)_L$ triplets breaks the $L \leftrightarrow R$ symmetry that is invoked in many left-right symmetric models~\cite{Duka:1999uc,Heeck:2015qra,Garcia-Cely:2015quu}. 

The triplets can be parametrized as:
\be
T_1 = \begin{pmatrix} t_1^+/\sqrt{2} & t_1^{++} \\ t_1^0 & -t_1^+/\sqrt{2} \end{pmatrix} ,~~ T_2 = \begin{pmatrix} t_2^-/\sqrt{2} & t_2^0 \\ t_2^{--} & -t_2^-/\sqrt{2} \end{pmatrix},
\ee
where the $0$ and $\pm$ superscripts are labels assigned with the foresight that these components will make up neutral or electrically charged fermions, accordingly (see Eq.~\ref{eq:charge}). The factors of $\sqrt{2}$ are fixed in order to guarantee canonical normalization of the kinetic terms. The most general renormalizable Lagrangian for the dark sector is given by:
\begin{align}
\mathcal{L} &\supset S^\dagger i \bar{\sigma}^\mu \partial_\mu S + \text{tr} (T_1^\dagger i \bar{\sigma}^\mu D_\mu T_1) + \text{tr} (T_2^\dagger i \bar{\sigma}^\mu D_\mu T_2)
\nl
&-\bigg[\frac{1}{2} M_S S^2 + M_T ~ \text{tr}(T_1 T_2) + \lambda_1 ~ S ~ \text{tr}(T_1 \Delta_R^\dagger) + \lambda_2 ~ S ~ \text{tr}(T_2 \Delta_R) + \text{h.c.}\bigg]
,
\label{singlettripletL}
\end{align}
where 2-component Weyl spinor indices are implied, and traces refer to sums over $SU(2)_R$ indices. $M_S$ and $M_T$ are the bare singlet and triplet masses, respectively, and $\lambda_{1,2}$ are dimensionless Yukawa couplings. After $\Delta_R$ acquires a VEV, these couplings generate mass terms for three Majorana fermions and two Dirac fermions,
\be
-\mathcal{L} \supset \frac{1}{2} \begin{pmatrix} S & t_1^0 & t_2^0 \end{pmatrix} \begin{pmatrix} M_S & \lambda_1 ~ v_R & \lambda_2 ~ v_R \\ \lambda_1 ~ v_R & 0 & M_T \\ \lambda_2 ~ v_R & M_T & 0 \end{pmatrix} \begin{pmatrix} S \\ t_1^0 \\ t_2^0 \end{pmatrix} + M_T t_1^+ t_2^- + M_T t_1^{++} t_2^{--} + \text{h.c.}
~,
\ee
where $S$ and $t_{1,2}^0$ denote the singlet and neutral triplet components, respectively. The electrically charged states are Dirac fermions of mass $M_T \, $:
\be
\x^+ \equiv \begin{pmatrix} t_1^+ \\ ( t_2^- )^\dagger \end{pmatrix} ~~~~~~\text{and}~~~~~~ \x^{++} \equiv \begin{pmatrix} t_1^{++} \\ ( t_2^{--} )^\dagger \end{pmatrix}
~.
\ee
Diagonalizing the neutral mass matrix yields the following decomposition:
\begin{align}
\label{eq:mixingangle}
S &= N_S^1 ~ \xI + N_S^2 ~ \xII + N_S^3 ~ \xIII
\nl
t_1^0 &= N_{t_1}^1 ~ \xI + N_{t_1}^2 ~ \xII + N_{t_1}^3 ~ \xIII
\nl
t_2^0 &= N_{t_2}^1 ~ \xI + N_{t_2}^2 ~ \xII + N_{t_2}^3 ~ \xIII
~.
\end{align}
Bearing in mind field redefinitions that fix wrong-sign mass terms such as $\x \to i \x \, $, the mixing angles in Eq.~(\ref{eq:mixingangle}) are promoted to complex numbers. Moving to 4-component notation, we define the Majorana spinors as follows: 
\be
\x_i^\text{(4-comp.)} \equiv \begin{pmatrix} \x_i  \\ \x_i^\dagger \end{pmatrix}.
\ee
Throughout the remainder of this paper, we will drop the superscripts for all 4-component spinors. In Appendix~\ref{appB}, we provide expressions describing the interactions between the singlet-triplet dark sector and gauge or Higgs bosons.

\begin{figure}[t]
\begin{center}
\includegraphics[width=0.45\textwidth]{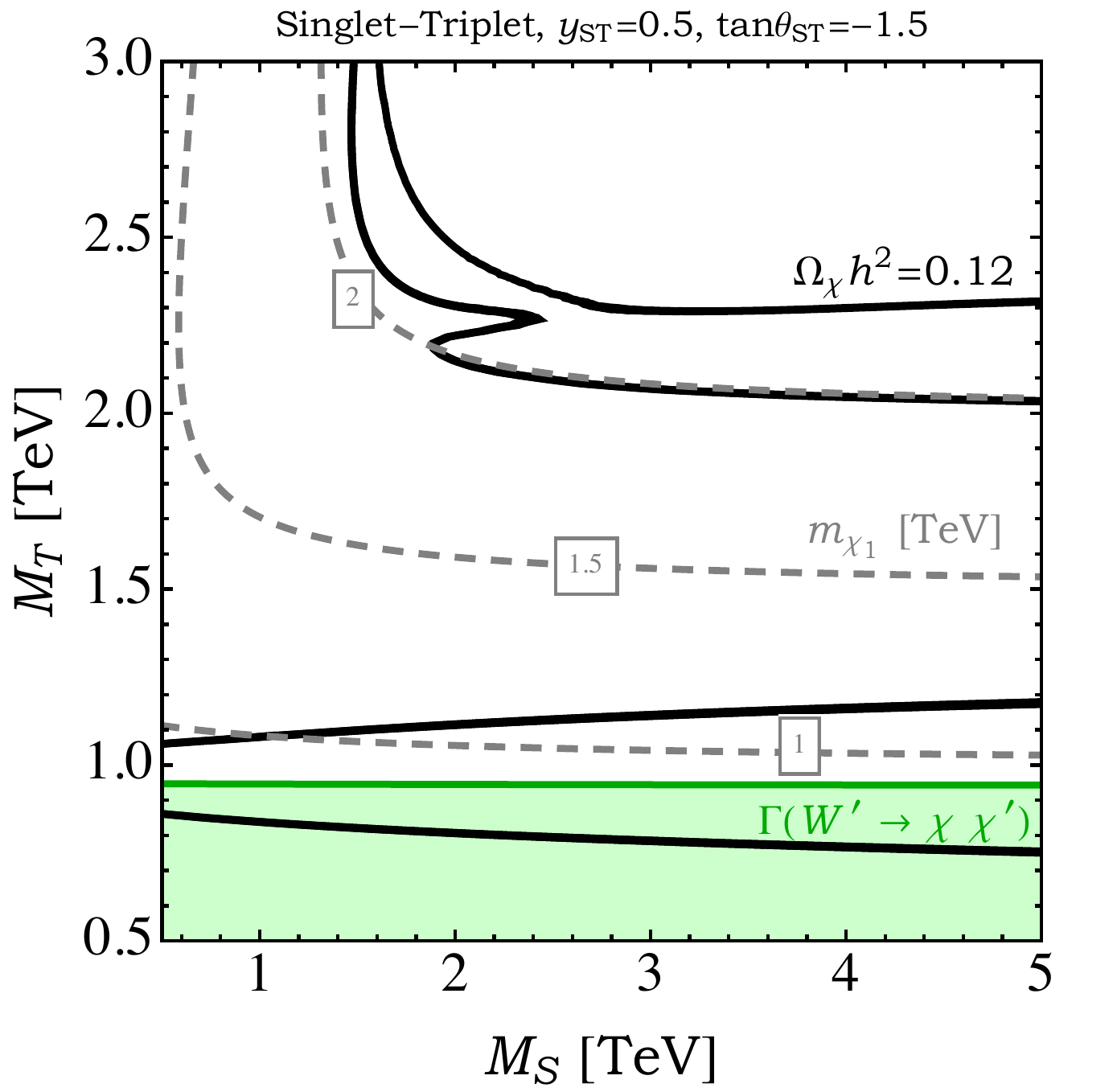} 
\includegraphics[width=0.45\textwidth]{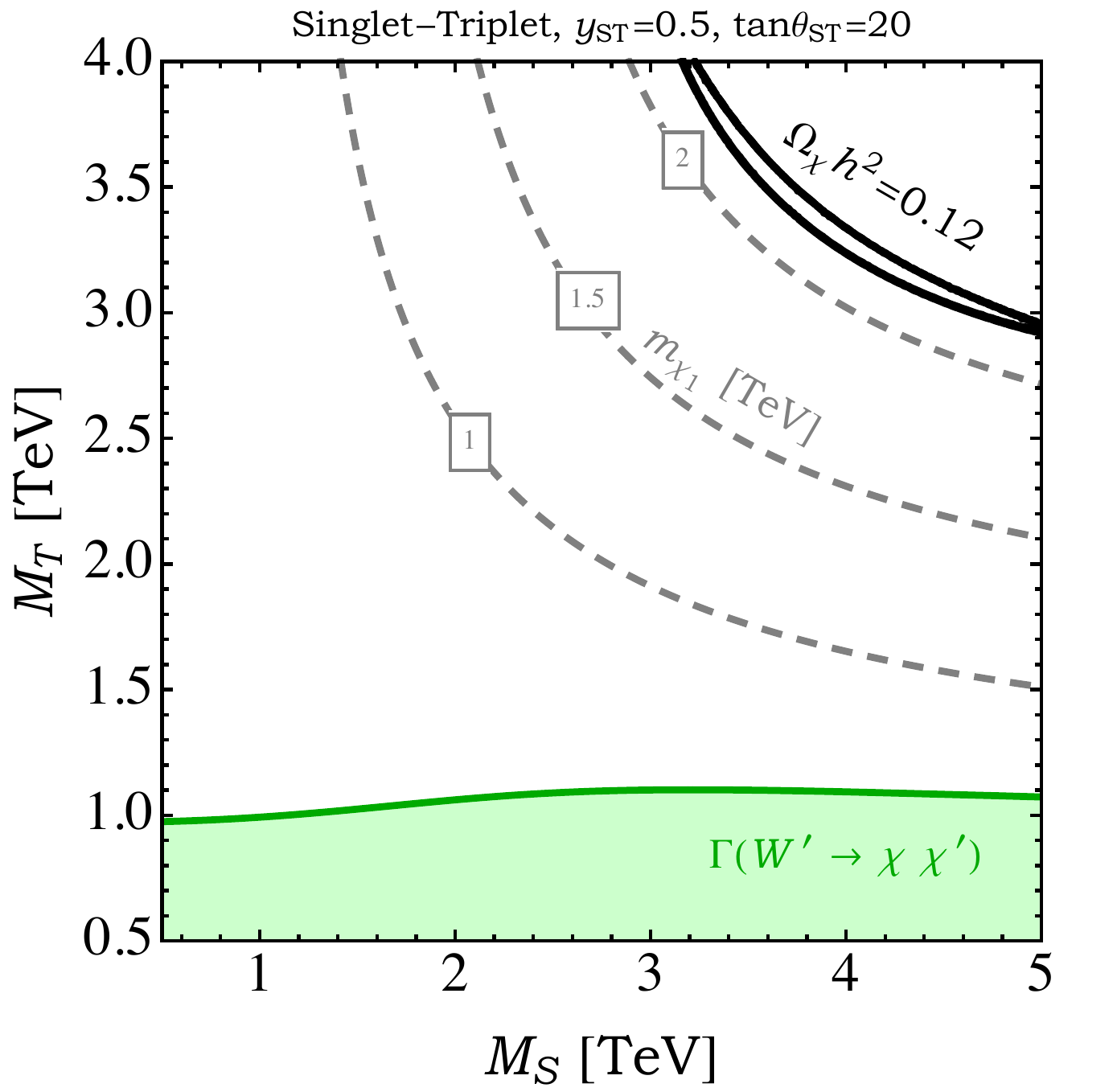} 
\caption{\label{fig:SingTrip1} Phenomenology of singlet-triplet dark matter. Along the solid black contours, the thermal relic abundance is in agreement with the measured cosmological dark matter density ($\Omega_{\chi} h^2=0.12)$. Also shown as dashed grey lines are contours of constant dark matter mass (as labeled). In this model, thermal freeze-out is dominated by resonant annihilation through the $Z'$ or resonant coannihilation through the $W'$. In each frame, we have adopted $g_R=0.45$ and $m_{W'} = 1.9$ TeV in order to match the rate and energy of the diboson excess, and $\tan \beta=2$ to accommodate the required $W^\prime \to W Z$ branching fraction. The green shaded regions are those in which the $W'$ decays to particles residing within the dark matter sector with a branching fraction greater than 10\%.}
\end{center}
\end{figure}

To calculate the thermal relic abundance of dark matter, we utilize the publicly available programs {\tt FeynRules}~\cite{Alloul:2013bka} and {\tt MicrOMEGAs}~\cite{Belanger:2013oya,Belanger:2014vza} and cross-check using {\tt MadDM}~\cite{Backovic:2015cra}. In Fig.~\ref{fig:SingTrip1}, we present examples of parameter space in which an abundance compatible with the measured cosmological dark matter density is obtained. We present these results in terms of the singlet and triplet masses, $M_S$ and $M_T$, and for various choices of the following parameters:
\begin{eqnarray}
y_{_{\rm ST}} &\equiv& \sqrt{\lambda_1^2 +\lambda_2^2 } ~ , \\
\tan &\theta_{_{\rm ST}}& \equiv \lambda_1/\lambda_2 ~ , \nonumber
\end{eqnarray}
where $\lambda_1$ and $\lambda_2$ are the Yukawa couplings introduced in Eq.~(\ref{singlettripletL}).  

The process of thermal freeze-out is largely governed by the masses of $\x_1$ and $\x^\pm$. In particular, the desired relic abundance is obtained near the $W^\prime$ or $Z^\prime$ resonances, corresponding to $m_{\x_1} \approx m_{W^\prime} / 2 \approx 1$ TeV or $m_{\x_1} \approx m_{Z^\prime}/2 \approx 2$ TeV, respectively. In the left frame of Fig.~\ref{fig:SingTrip1}, $|\tan{\theta_\text{ST}}| \sim 1$, in which case there is an enhanced parity symmetry acting on the triplets $T_{1,2}$. As a result, the singlet, $S$, mixes with only one linear combination of $t_1^0$ and $t_2^0$, while the other remains degenerate with the charged states. For sufficiently small values of the triplet mass, $M_T \lesssim M_S$, the lightest neutral and charged fermions in the dark sector are approximately degenerate, leading to efficient coannihilations in the early universe through the $s$-channel exchange of a $W^\prime$. Alternatively, in the right frame of this figure, $|\tan{\theta_\text{ST}}| \gg 1$, and there is a significant mass splitting between the neutral and charged states, suppressing the role of coannihilations. In each frame, we have adopted $g_R = 0.45$, $m_{W^\prime} = 1.9$ TeV, and $\tan{\beta} = 2$, motivated by the observed characteristics of the diboson excess. We note that if both the neutral component of the Higgs triplet, $\Delta_0$, and the right-handed neutrinos, $\nu_R$, are relatively light, dark matter annihilations to a $\nu_R \nu_R$ final state could play a significant role in the determination of the relic abundance. Throughout this paper, however, we will assume that these states are heavy and neglect their contribution.

In this model, the elastic scattering of dark matter with nuclei is dominated by $Z^\prime$ exchange. The cross section for this process is spin-dependent, and of the following magnitude:
\be
\sigma_\text{SD} \approx 2 \times 10^{-45} \, \text{cm}^2 \, \times \left( \frac{g_{Z^\prime}^{(1)}}{0.1}\right)^2 \left( \frac{4 \text{ TeV}}{m_{Z^\prime}}\right)^4,
\ee
where $g_{Z^{\prime}}^{(1)}$ is the dark matter's coupling to the $Z'$ (see Appendix~\ref{appB}). Even for relatively large values of this coupling (corresponding to a large value of $|N^1_{t_1}|^2-|N^1_{t_2}|^2$), the predicted cross section is well below the reach of current and planned experiments, and likely below the so-called ``neutrino floor''~\cite{Billard:2013qya}.

Also shown in Fig.~\ref{fig:SingTrip1}, are the regions of parameter space in which the $W^\prime$ has a large branching fraction to the dark sector. This is motivated by the fact that the rate associated with the diboson excess naively requires $g_R \approx 0.4 - 0.6$, which is in contrast to some theoretical expectations favoring $g_R = g_L \approx 0.65$. Decays of the $W^\prime$ to particles in the dark sector could plausibly accommodate such an equality. We estimate that this would require a branching fraction of several tens of percents. However, given the rough nature of this estimate, in Fig.~\ref{fig:SingTrip1} (as well as in Figs.~\ref{fig:SingBid1}~and~\ref{fig:TripBid1}), we show, for illustration, regions of parameter space where $\text{BR}(W^\prime \to \text{dark sector}) \gsim 10 \% \, $. 

In order to ensure a viable dark matter candidate, it is imperative that the lightest dark sector state is electrically neutral. In the decoupled limit, $M_T \ll M_S$, the lightest neutral and charged states are nearly degenerate at tree-level, suggesting that radiative corrections are potentially important. As described in Appendix~\ref{sec:loops}, we calculate the full set of one-loop corrections to the dark sector masses and find that $m_{\x^\pm}, m_{\x^{\pm \pm}} > m_{\x_1}$ throughout the entirety of the parameter space shown in Fig.~\ref{fig:SingTrip1}. Alternatively, for $ M_S \gtrsim 50 \text{ TeV} \gg M_T$, radiative corrections lead to $m_{\x^\pm} , m_{\x^{\pm \pm}} < m_{\x_1}$ when $\tan{\theta_\text{ST}} =-1.5$ and the remaining parameters are chosen as in Fig.~\ref{fig:SingTrip1}.

\subsection{Singlet-Bidoublet Dark Matter}
\label{sec:singletbidoublet}

In this section, we introduce two Weyl fermions: a singlet, $S$, and a bidoublet, $B$, with charge assignments as given in Table~\ref{table:singletbidoublet}. The bidoublet is parametrized as follows:
\be
B = \begin{pmatrix} b_1^0 & -b_2^+ \\ b_1^- & b_2^0 \end{pmatrix},
\ee
where the $0$ and $\pm$ superscripts are labels chosen with the foresight that these fermions will make up neutral or electrically charged fermions, accordingly (see Eq.~\ref{eq:charge}). The most general renormalizable Lagrangian for the dark sector is given by:
\begin{align}
\mathcal{L} &\supset S^\dagger i \bar{\sigma}^\mu \partial_\mu S + \text{tr} (B^\dagger i \bar{\sigma}^\mu D_\mu B)
\nl
&-\bigg[ \frac{1}{2} M_S S^2 + \frac{1}{2} M_B ~ \text{tr}(B \tilde{B}^\dagger) + \lambda ~ S ~ \text{tr}(B \phi^\dagger) + \tilde{\lambda} ~ S ~ \text{tr}(B \tilde{\phi}^\dagger) + \text{h.c.}\bigg],
\label{singletbidoubletL}
\end{align}
where 2-component Weyl spinor indices are implied, and traces refer to sums over $SU(2)$ indices. $M_S$ and $M_B$ denote the bare singlet and bidoublet masses, respectively, and $\lambda$ and $\tilde{\lambda}$ are dimensionless Yukawa couplings. After EWSB, the dark sector fermion masses are described by,
\be
-\mathcal{L} \supset \frac{1}{2} \begin{pmatrix} S & b_1^0 & b_2^0 \end{pmatrix} \begin{pmatrix} M_S & v (\lambda c_\beta + \tilde{\lambda} s_\beta) & v (\lambda s_\beta + \tilde{\lambda} c_\beta) \\ v (\lambda c_\beta + \tilde{\lambda} s_\beta) & 0 & M_B \\ v (\lambda s_\beta + \tilde{\lambda} c_\beta) & M_B & 0 \end{pmatrix} \begin{pmatrix} S \\ b_1^0 \\ b_2^0 \end{pmatrix} + M_B b_1^- b_2^+ + \text{h.c.}~.
\ee
Whereas $S$ and the neutral bidoublet components, $b_{1,2}^0$, mix to form three Majorana fermions, the charged components constitute a single charged Dirac fermion of mass $M_B$:
\be
\x^+ \equiv \begin{pmatrix} b_2^+ \\ ( b_1^- )^\dagger \end{pmatrix}.
\ee
\begin{table}[t]
\centering
\begin{tabular}{| c || c | c | c |}
\hline
Field & Charges & Spin \\ \hline \hline
$S$ & $({\bf 1},{\bf 1},0)$ & 1/2 \\ \hline
$B$ & $({\bf 2},{\bf 2},0)$ & 1/2 \\ \hline
\end{tabular}
\caption{The $SU(2)_L$, $SU(2)_R$, and $B-L$ charge assignments in the singlet-bidoublet model. All fields are colorless.}
\label{table:singletbidoublet}
\end{table}
As before, we diagonalize the neutral mass matrix by decomposing the neutral gauge eigenstates in terms of the mass eigenstates as,
\begin{align}
\label{eq:mixingangle2}
S &= N_S^1 ~ \xI + N_S^2 ~ \xII + N_S^3 ~ \xIII
\nl
b_1^0 &= N_{b_1}^1 ~ \xI + N_{b_1}^2 ~ \xII + N_{b_1}^3 ~ \xIII
\nl
b_2^0 &= N_{b_2}^1 ~ \xI + N_{b_2}^2 ~ \xII + N_{b_2}^3 ~ \xIII~,
\end{align}
and we again adopt 4-component notation for the Majorana spinors. In Appendix~\ref{appC}, we provide expressions describing the interactions between the singlet-bidoublet dark sector and gauge or Higgs bosons.

\begin{figure}[t]
\begin{center}
\includegraphics[width=0.45\textwidth]{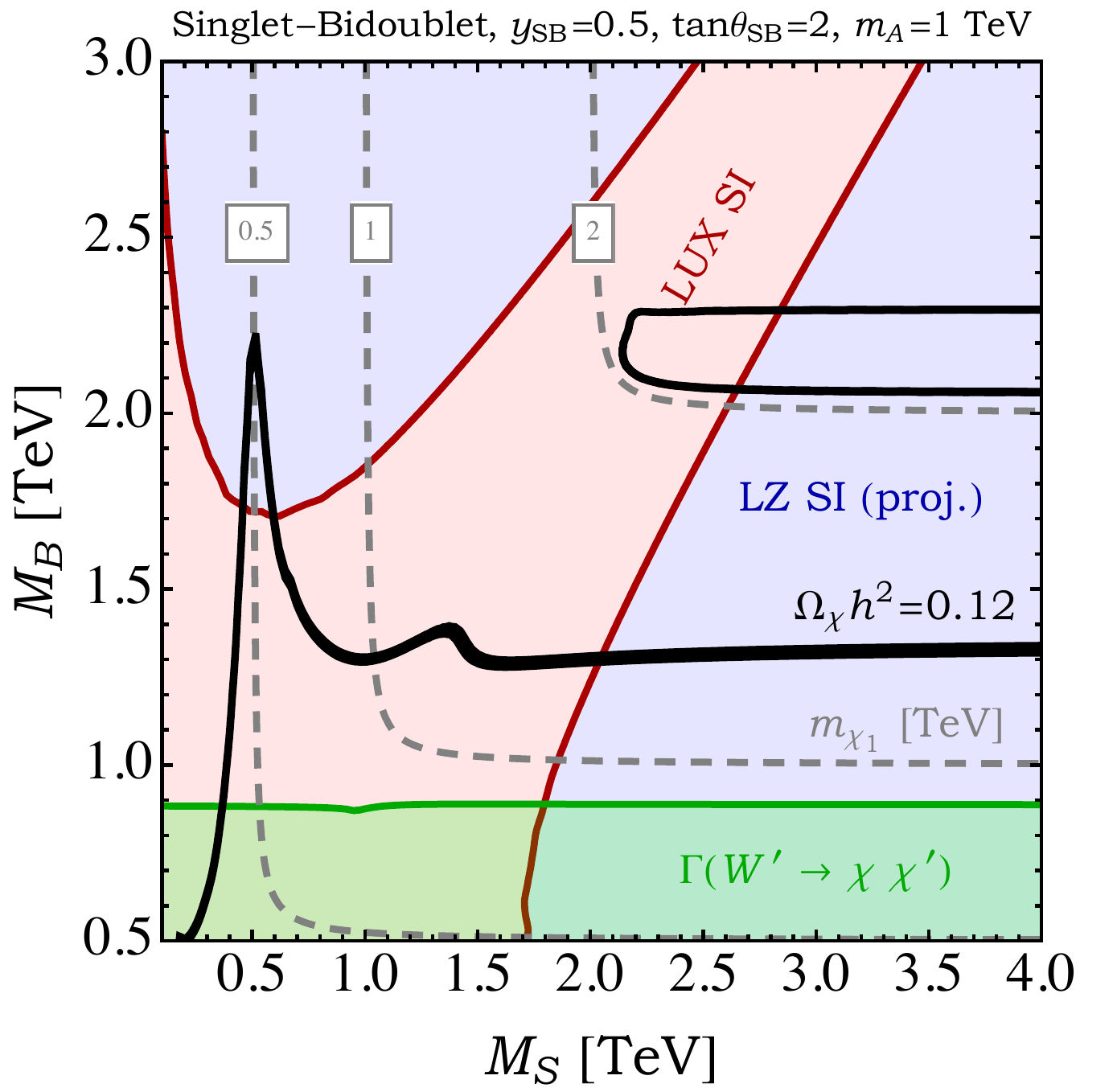} 
\includegraphics[width=0.45\textwidth]{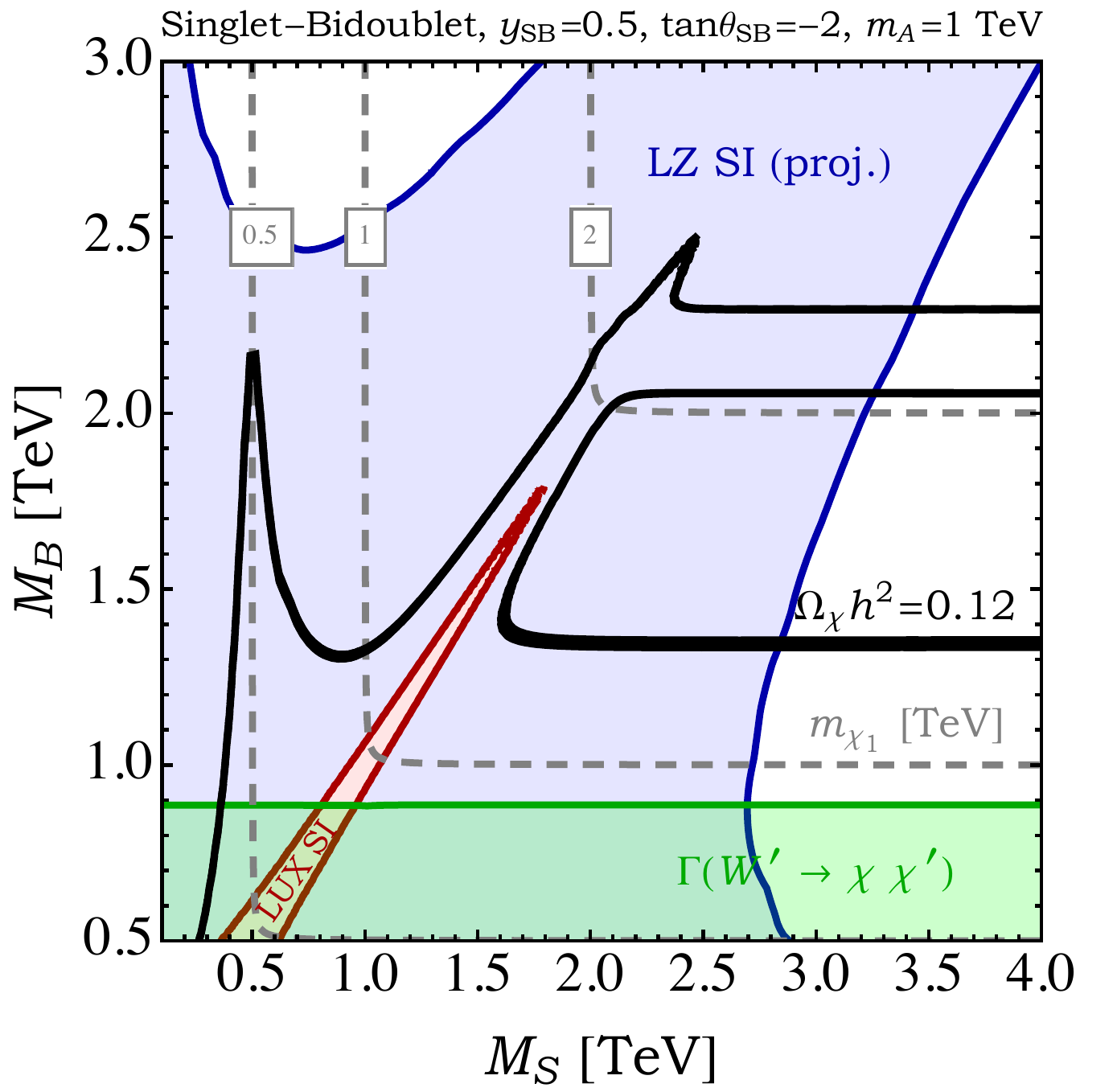} 
\includegraphics[width=0.45\textwidth]{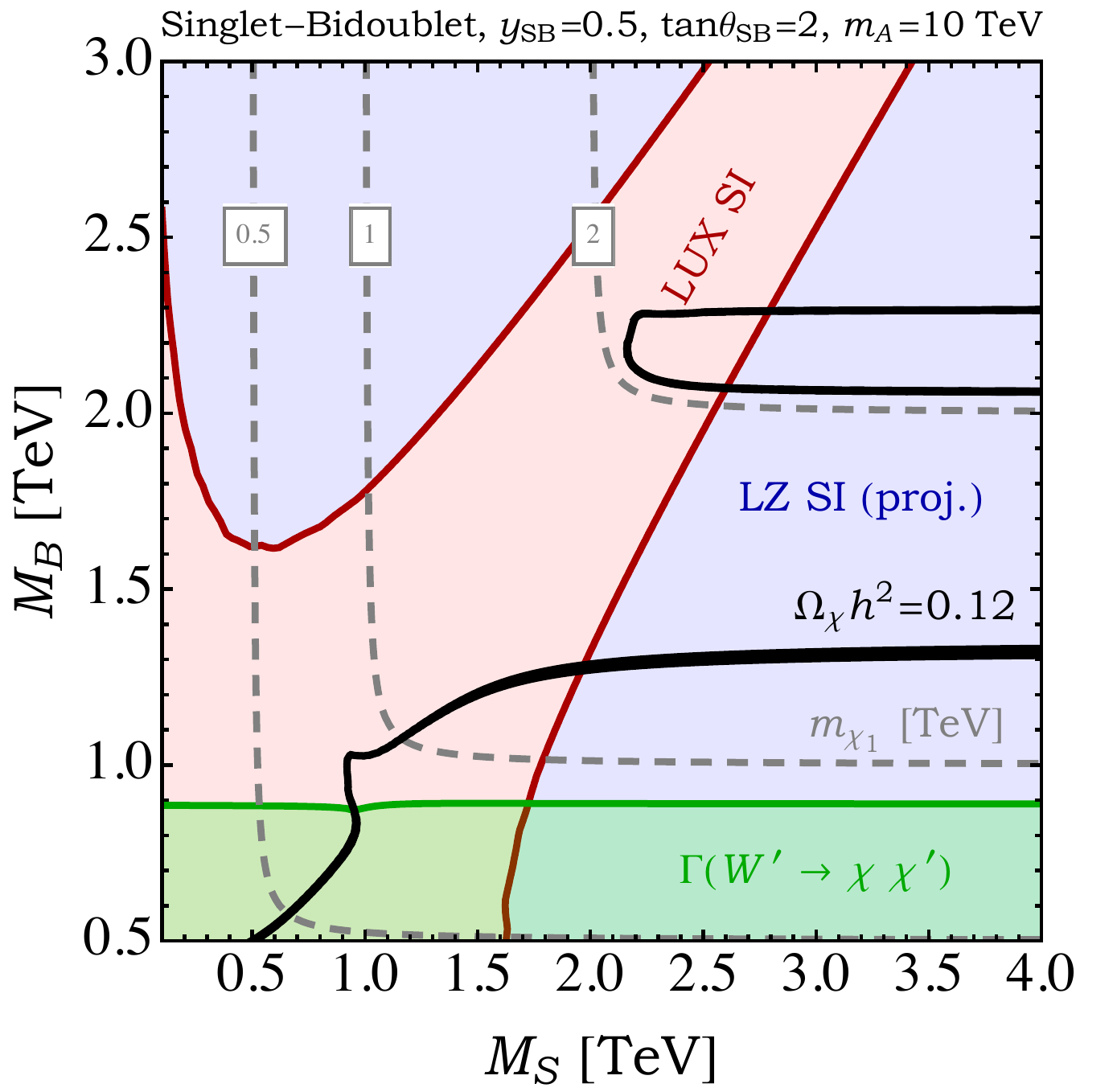}
\includegraphics[width=0.45\textwidth]{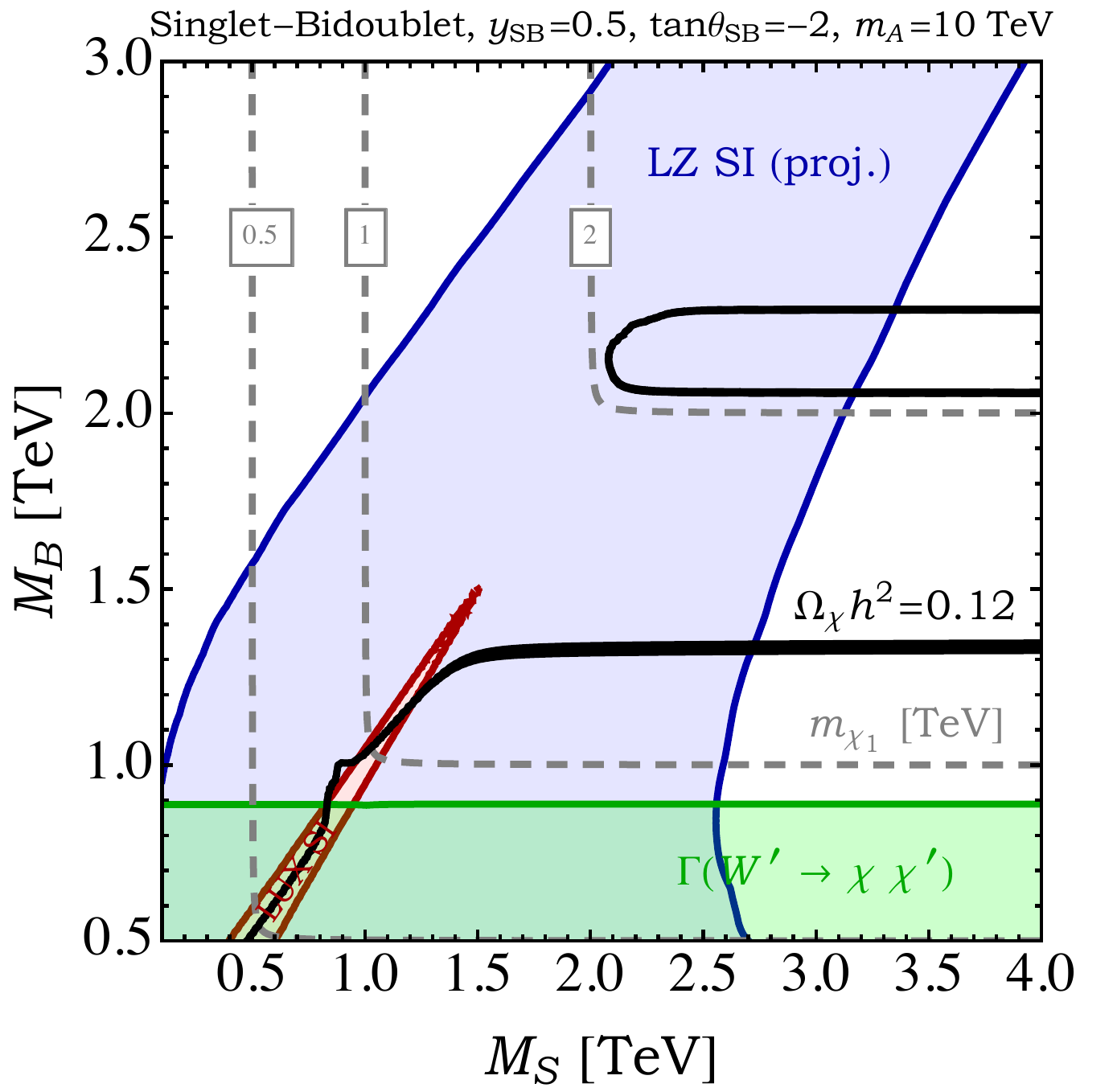}
\caption{\label{fig:SingBid1} Phenomenology of singlet-bidoublet dark matter. Along the solid black contours, the thermal relic abundance is in agreement with the measured cosmological dark matter density ($\Omega_{\chi} h^2=0.12)$. Also shown as dashed grey lines are contours of constant dark matter mass (as labeled). In each frame, we have adopted $g_R=0.45$ and $m_{W'} = 1.9$ TeV in order to match the rate and energy of the diboson excess, and $\tan \beta=2$ to accommodate the required $W^\prime \to W Z$ branching fraction. The red shaded regions are currently excluded by LUX, whereas the blue regions are predicted to fall within the reach of LZ. The green shaded regions are those in which the $W'$ decays to particles residing within the dark matter sector with a branching fraction greater than 10\%.}
\end{center}
\end{figure}

In Fig.~\ref{fig:SingBid1}, we explore some of the phenomenological features of this model, presenting our results in terms of  $M_B$, $M_S$, $m_A$ ($=m_H, m_{H^{\pm}}$), and the parameters: 
\begin{eqnarray}
y_{_{\rm SB}} &\equiv& \sqrt{\lambda_{ }^2+\tilde{\lambda}^2}, \\
\tan \theta_{_{\rm SB}} &\equiv& \lambda\,/\,\tilde{\lambda}\,, \nonumber
\end{eqnarray}
where $\lambda$ and $\tilde{\lambda}$ are the Yukawa couplings as defined in Eq.~(\ref{singletbidoubletL}).  

Dark matter freeze-out is largely dictated by annihilations and coannihilations through the $s$-channel exchange of a $Z^\prime$ or $W^\prime$ gauge boson. Additional annihilation channels become active if the heavy Higgs bosons have masses that are comparable to $m_{\x_1}$, in which case a region of parameter space analogous to the $A$-funnel in the MSSM is found near $m_{\x_1} \approx m_A / 2$. This is related to our choice of Yukawa structure in Eq.~(\ref{eq:yukawas}). In regions of parameter space with a light and mostly singlet dark matter candidate and a relatively light pseudoscalar Higgs, it may be possible to generate the Galactic Center gamma-ray excess~\cite{Goodenough:2009gk,Hooper:2010mq,Hooper:2011ti,Daylan:2014rsa,Calore:2014xka,TheFermi-LAT:2015kwa} in this model, similar to as in the models described in Refs.~\cite{Ipek:2014gua,Berlin:2015wwa}. Furthermore, depending on the sign of $\tan{\theta}$, singlet mixing allows for enhanced annihilations through heavy scalars when $m_{\x_1} \sim M_S \sim M_B$. If, on the other hand, the heavy Higgses are decoupled, proper freeze-out favors regions where $\x_1$ is predominantly bidoublet-like, and annihilations involving heavy gauge bosons lead to the correct relic density near the $W^\prime$ and $Z^\prime$ resonances.

For simplicity, we have ignored trilinear Higgs interactions involving one or more heavy scalar since they depend explicitly on the $\sim 10$ parameters of the general Higgs potential. As a result, we have purposely neglected annihilation processes such as $\chi \chi \rightarrow A \rightarrow A h$ in the evaluation of the dark matter relic density. However, since we do not consider dark matter masses much greater than a few TeV, we do not expect these interactions to dominate in any of the parameter space shown.

Elastic scattering between dark matter and nuclei is dominated by SM Higgs exchange, leading to a spin-independent cross section that may be within the reach of current or future direct detection experiments. The cross section for this process is spin-independent, and of the following magnitude:
\be
\sigma_\text{SI} \approx 2 \times 10^{-44} \, \text{cm}^2 \, \times \left( \frac{\lambda_h^{(1)}}{0.1}\right)^2,
\label{SIhiggs}
\ee
where $\lambda_{h}^{(1)}$ is the dark matter couplings to the light Higgs (see Appendix~\ref{appC}). In Fig.~\ref{fig:SingBid1}, the shaded red regions are currently excluded by the constraints from LUX~\cite{Akerib:2015rjg}, whereas the shaded blue regions fall within the projected reach of LZ \cite{Cushman:2013zza}. In calculating the dark matter coupling to nucleons, we have taken the scalar nucleon form factors as listed in Sec.~4 of Ref.~\cite{Hill:2014yxa}.

Dark matter-nucleon scattering is suppressed for negative values of $\tan\theta_{{\rm SB}}$.  To understand this behavior, note that the dark matter-Higgs coupling can be determined, from low energy theorems, to be proportional to $\frac{\partial m_{\x_1}}{\partial v}\, $, which in turn can be found by differentiating the characteristic equation, $\det \left(M_N - m_{\x_1} \mathbb{1} \right) = 0$,
with respect to the Higgs VEV \cite{Cohen:2011ec,Cheung:2012qy}.  We find that the dark matter-Higgs coupling scales as,
\be
\lambda_h^{(1)} \propto ( 1 + \sin 2 \beta \sin 2 \theta_{_{\rm SB}} ) m_{\x_1} + ( \sin 2 \beta + \sin 2 \theta_{_{\rm SB}} ) M_B~.
\ee
Note that there is a suppression of $\lambda_h^{(1)}$ if $\tan\beta$ and $\tan\theta_{_{\rm SB}}$ are of opposite sign, as can be seen in the right-hand plots of Fig.~{\ref{fig:SingBid1}}.  Furthermore, in this regime, there may be a direct detection blind spot (corresponding to a vanishing elastic scattering cross section) for non-trivial mixing when $\lambda_h^{(1)}=0$, or equivalently  
\be
m_{\x_1} = -\frac{ \sin 2 \beta + \sin 2 \theta_{_{\rm SB}} }{1 + \sin 2 \beta \sin 2 \theta_{_{\rm SB}}} M_B
~.
\ee

In the limit that $M_B \ll M_S$, the lightest neutral and charged states are nearly degenerate at tree-level. In this case, it is important to investigate whether radiative corrections guarantee that the lightest dark sector state is electrically neutral, crucial for any dark matter candidate. As outlined in Appendix~\ref{sec:loops}, we calculate the full set of leading order radiative corrections to the masses of the lightest neutral and charged states in the dark sector. At the one-loop level, $m_{\x^\pm} > m_{\x_1}$ throughout the parameter space shown in Fig.~\ref{fig:SingBid1}. However, in the case that $M_B = 500 $ GeV, $y_{\text{SB}} = 0.5$, and $\tan{\theta_\text{SB}} = -2$, we find that $m_{\x^\pm} < m_{\x_1}$ for singlet masses as large as $M_S \gtrsim 50$ TeV.

\subsection{Triplet-Bidoublet Dark Matter}
\label{sec:tripletbidoublet}

In this model, we introduce two Weyl fermions: a triplet, $T$, and a bidoublet, $B$, with charges as shown in Table~\ref{table:tripletbidoublet}. The triplet and bidoublet are parametrized as:
\be
T = \begin{pmatrix}  t^0/\sqrt{2} & t_2^+ \\ t_1^- & -t^0/\sqrt{2} \end{pmatrix} ,~~~ B = \begin{pmatrix}  b_1^0 & -b_2^+ \\ b_1^- & b_2^0 \end{pmatrix},
\ee
where the $0$ and $\pm$ superscripts are labels assigned with the foresight that these fermions will make up neutral or electrically charged fermions, accordingly (see Eq.~\ref{eq:charge}), and the factors of $\sqrt{2}$ are fixed in order to guarantee canonical normalization of the kinetic terms. The most general renormalizable Lagrangian for the dark sector is given by:
\begin{eqnarray}
\mathcal{L} &\supset&  \text{tr} (T^\dagger i \bar{\sigma}^\mu D_\mu T) + \text{tr} (B^\dagger i \bar{\sigma}^\mu D_\mu B)
\nl
&-&\bigg[ \frac{1}{2} M_T ~ \text{tr}(T^2) + \frac{1}{2} M_B ~ \text{tr}(B \tilde{B}^\dagger) + \lambda ~ \text{tr}(B T \phi^\dagger) + \tilde{\lambda} ~ \text{tr}(B T \tilde{\phi}^\dagger) + \text{h.c.}\bigg],
\label{TBkin}
\end{eqnarray}
where 2-component Weyl spinor indices are implied, and traces refer to sums over $SU(2)$ indices. $M_T$ and $M_B$ are bare triplet and bidoublet masses, respectively, and $\lambda$ and $\tilde{\lambda}$ are dimensionless Yukawa couplings. After EWSB, the neutral and charged fermions mix according to the following mass matrices:
\begin{eqnarray}
-\mathcal{L} &\supset& \frac{1}{2} \begin{pmatrix} t^0 & b_1^0 & b_2^0 \end{pmatrix} \begin{pmatrix} M_T & v (\lambda c_\beta + \tilde{\lambda} s_\beta) / \sqrt{2} & -v (\lambda s_\beta + \tilde{\lambda} c_\beta) / \sqrt{2}  \\  v (\lambda c_\beta + \tilde{\lambda} s_\beta) / \sqrt{2} & 0 & M_B \\  -v (\lambda s_\beta + \tilde{\lambda} c_\beta) / \sqrt{2} & M_B & 0 \end{pmatrix} \begin{pmatrix} t^0 \\ b_1^0 \\ b_2^0 \end{pmatrix}
\nl
&+& \begin{pmatrix} t_2^+ & b_2^+ \end{pmatrix} \begin{pmatrix} M_T & v (\lambda s_\beta + \tilde{\lambda} c_\beta) \\ - v (\lambda c_\beta + \tilde{\lambda} s_\beta) & M_B \end{pmatrix} \begin{pmatrix} t_1^- \\ b_1^- \end{pmatrix} + \text{h.c.} ~.
\label{cmm}
\end{eqnarray}
The neutral triplet, $t^0$, and the neutral bidoublet components, $b_{1,2}^0$, mix to form three Majorana fermions, while the charged components mix to form two charged Dirac fermions. Diagonalizing the neutral mass matrix yields the following decomposition:
\begin{align}
\label{eq:mixingangle3}
t^0 &= N_t^1 ~ \xI + N_t^2 ~ \xII + N_t^3 ~ \xIII
\nl
b_1^0 &= N_{b_1}^1 ~ \xI + N_{b_1}^2 ~ \xII + N_{b_1}^3 ~ \xIII
\nl
b_2^0 &= N_{b_2}^1 ~ \xI + N_{b_2}^2 ~ \xII + N_{b_2}^3 ~ \xIII
~.
\end{align}
\begin{table}[t]
\centering
\begin{tabular}{| c || c | c | c |}
\hline
Field & Charges & Spin \\ \hline \hline
$T$ & $({\bf 1},{\bf 3},0)$ & 1/2 \\ \hline
$B$ & $({\bf 2},{\bf 2},0)$ & 1/2 \\ \hline
\end{tabular}
\caption{The $SU(2)_L$, $SU(2)_R$, and $B-L$ charge assignments in the triplet-bidoublet model. All fields are colorless.}
\label{table:tripletbidoublet}
\end{table}
Bearing in mind field redefinitions that fix wrong-sign mass terms such as $\x \to i \x$, the mixing angles in Eq.~(\ref{eq:mixingangle3}) are promoted to complex numbers. Similarly, for the charged states:
\begin{align}
t_1^- &= U_{11} ~ \xm_1 + U_{12} ~ \xm_2
\nl
b_1^- &= U_{21} ~ \xm_1 + U_{22} ~ \xm_2
\nl
t_2^+ &= V_{11} ~ \xp_1 + V_{12} ~ \xp_2
\nl
b_2^+ &= V_{21} ~ \xp_1 + V_{22} ~ \xp_2
~.
\end{align}
Above, $U_{ij}$ and $V_{ij}$ are orthogonal matrices that are constructed from the eigenvectors of ${\bf M}^{\dagger} {\bf M}$ and ${\bf M} {\bf M}^{\dagger}$, respectively, where ${\bf M}$ is the charged mass matrix of Eq.~(\ref{cmm}). Once again, we adopt 4-component notation for the Majorana and Dirac spinors. In Appendix~\ref{appD}, we provide expressions describing the interactions between the triplet-bidoublet dark sector and gauge or Higgs bosons.

\begin{figure}[t]
\begin{center}
\includegraphics[width=0.45\textwidth]{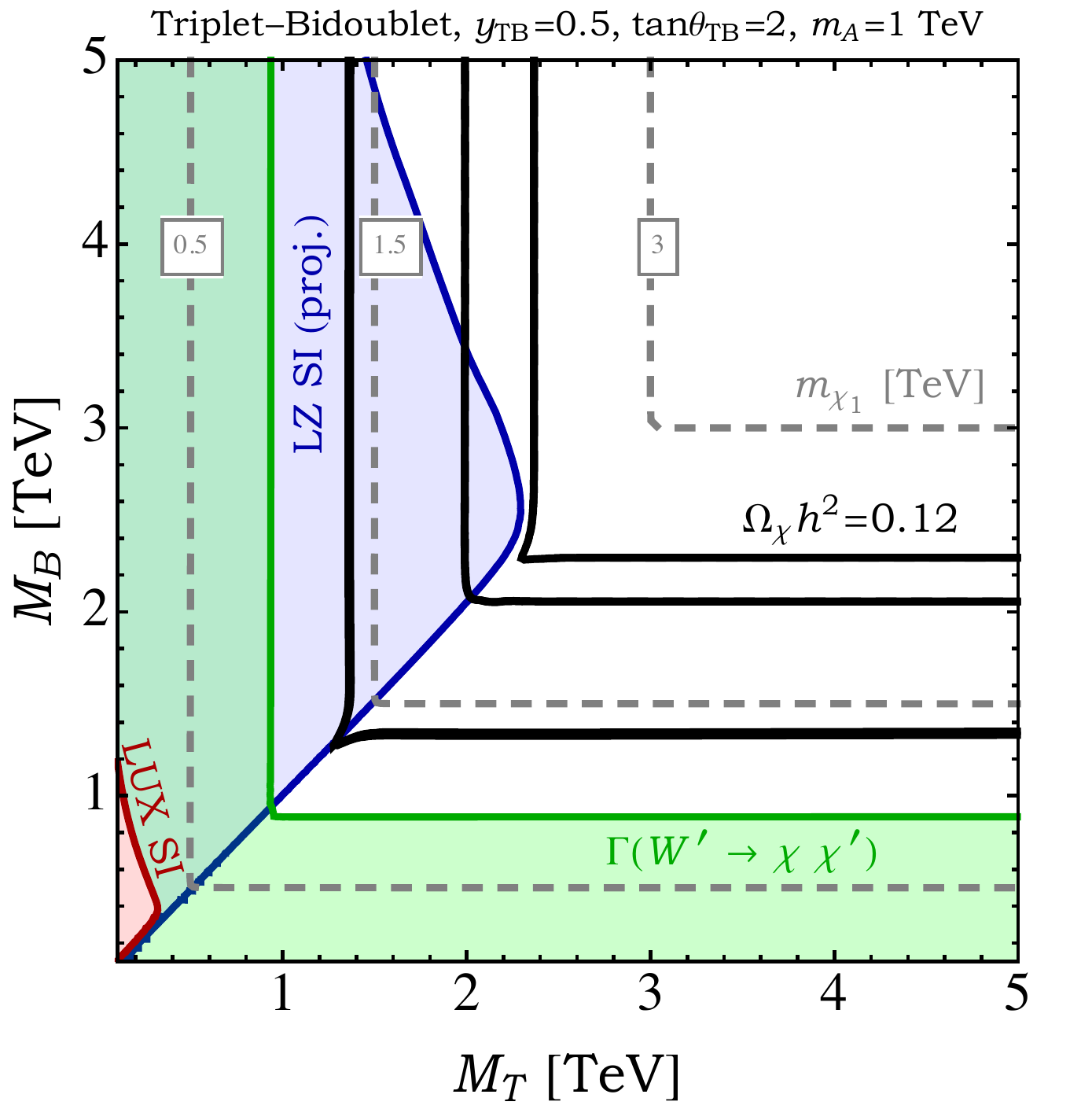}
\includegraphics[width=0.45\textwidth]{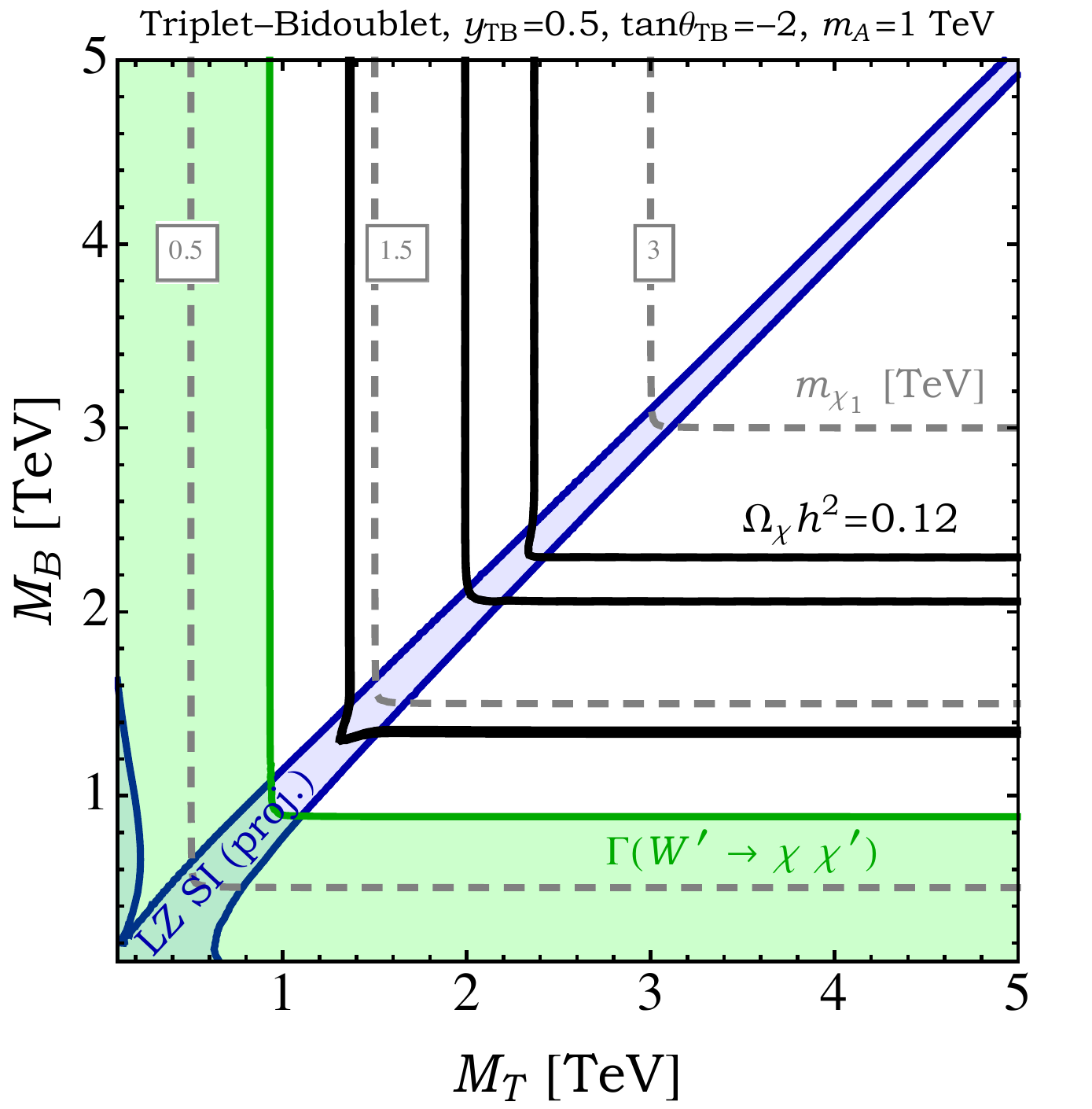}\\
\includegraphics[width=0.45\textwidth]{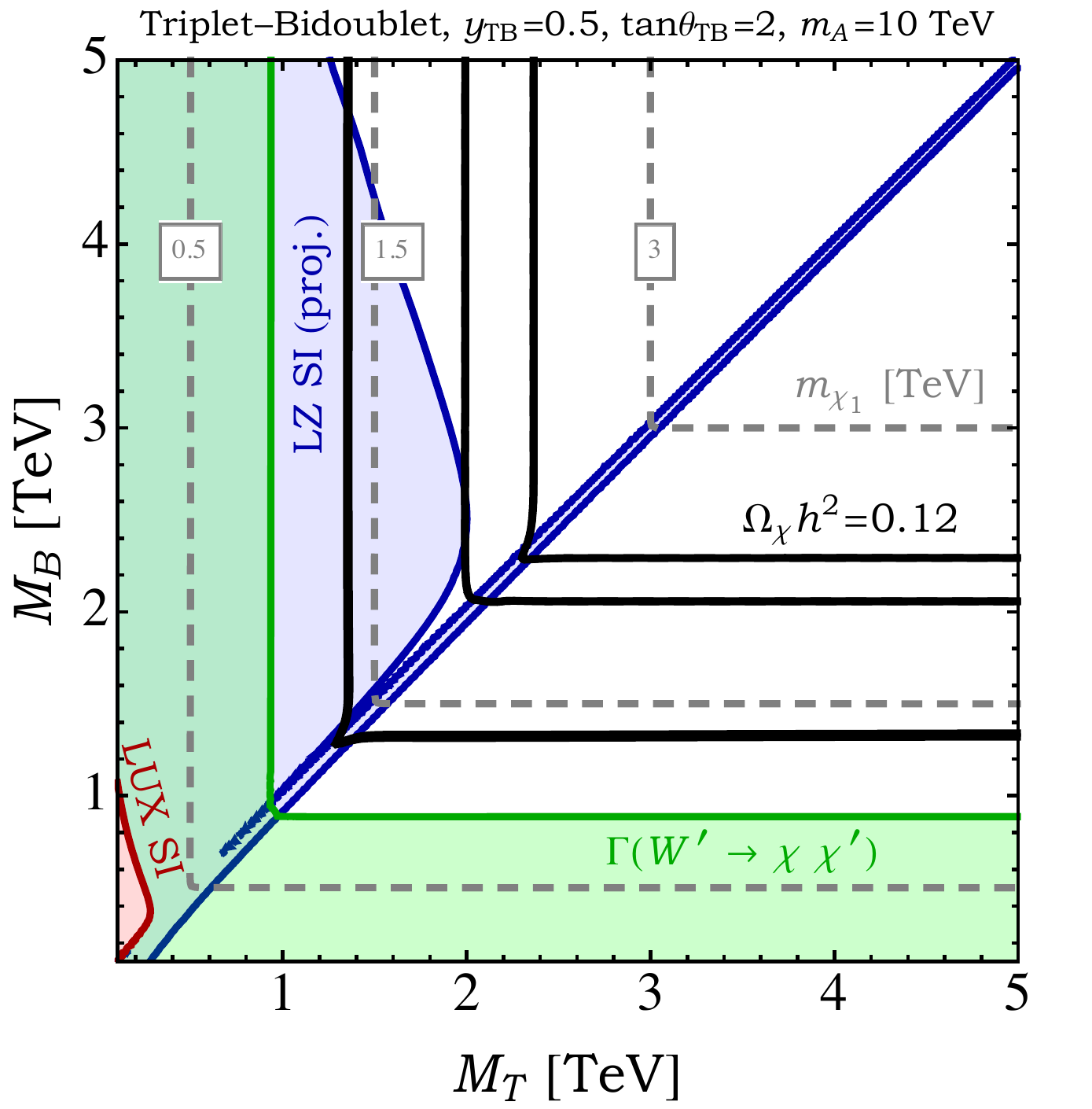} 
\includegraphics[width=0.45\textwidth]{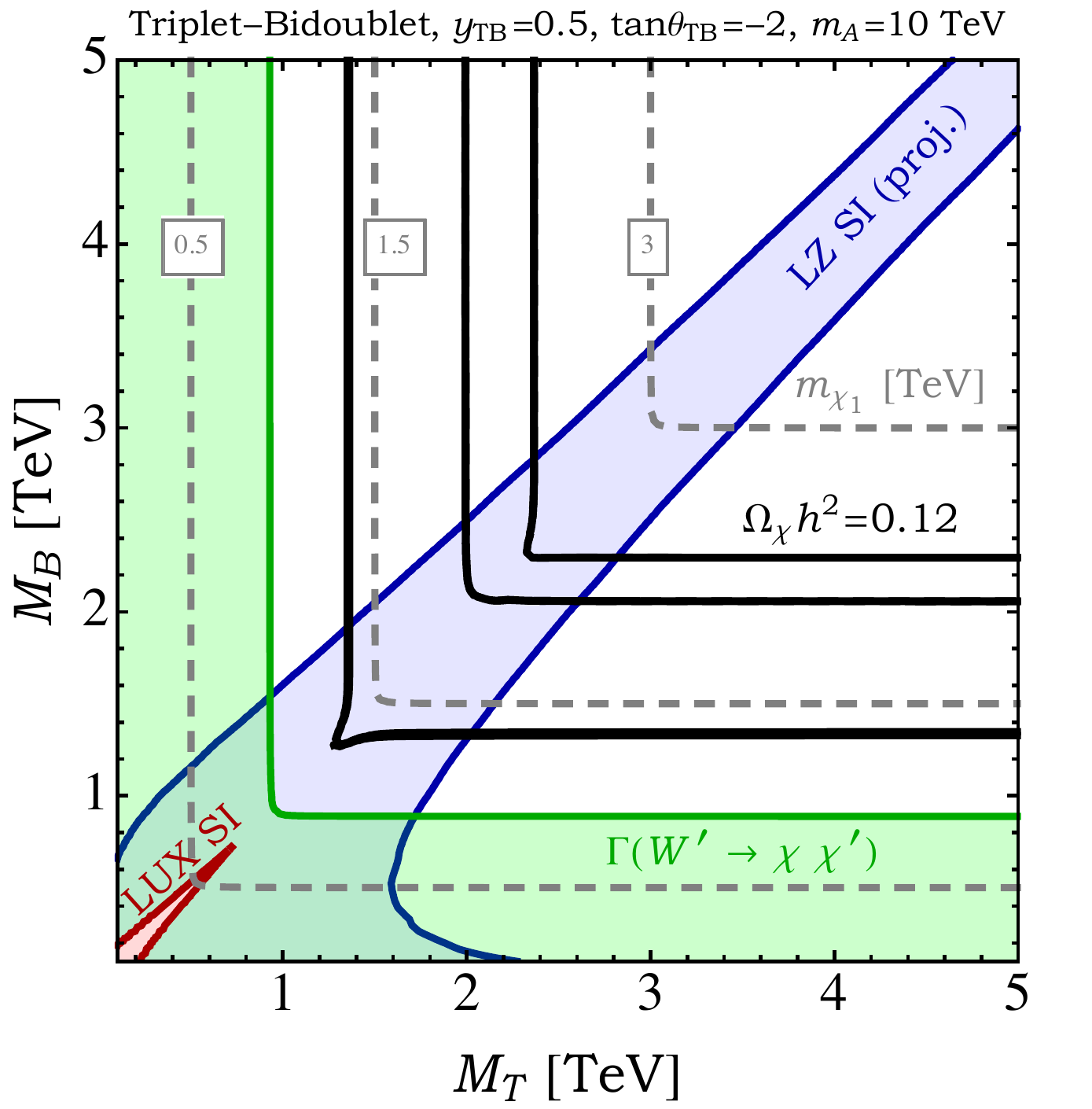} 
\caption{\label{fig:TripBid1} Phenomenology of triplet-bidoublet dark matter. Along the solid black contours, the thermal relic abundance is in agreement with the measured cosmological dark matter density ($\Omega_{\chi} h^2=0.12)$. Also shown as dashed grey lines are contours of constant dark matter mass (as labeled). In each frame, we have adopted $g_R=0.45$ and $m_{W'} = 1.9$ TeV in order to match the rate and energy of the diboson excess, and $\tan \beta=2$ to accommodate the required $W^\prime \to W Z$ branching fraction. The red shaded regions are currently excluded by LUX, whereas the blue regions are predicted to fall within the reach of LZ. The green shaded regions are those in which the $W'$ decays to particles residing within the dark matter sector with a branching fraction greater than 10\%.}
\end{center}
\end{figure}

In Fig.~\ref{fig:TripBid1}, we present some of the phenomenological features of this model, describing the parameter space in terms of  $M_B$, $M_T$, $m_A$ ($=m_H, m_{H^{\pm}}$), and the following: 
\begin{eqnarray}
y_{_{\rm TB}} &\equiv& \sqrt{\lambda_{ }^2+\tilde{\lambda}^2}, \\
\tan \theta_{_{\rm TB}} &\equiv& \lambda\,/\,\tilde{\lambda}\,, \nonumber
\end{eqnarray}
where $\lambda$ and $\tilde{\lambda}$ are the Yukawa couplings as defined in Eq.~(\ref{TBkin}).  

Similar to the singlet-bidoublet case described in the previous subsection, dark matter freeze-out is governed primarily by annihilation either through $s$-channel $W^\prime$ or $Z^\prime$ exchange. 
The elastic scattering between dark matter and nuclei is also dominated by SM Higgs exchange, with a cross section that is the same as given in Eq.~(\ref{SIhiggs}) (but using the expression for $\lambda_h^{(1)}$ found in Appendix~\ref{appD}). At present, the limits from LUX~\cite{Akerib:2015rjg} exclude only a very small portion of the otherwise viable parameter space in this model (red shaded), although the future reach of experiments such as LZ is projected to be much more expansive (shaded blue).  Note that heavy Higgs exchange destructively interferes with SM Higgs exchange and slightly suppresses the direct detection rate for $M_B\approx M_T$, which is most noticeable for $\tan\theta_{_{\rm TB}} > 0$. From the low energy theorem, the dark matter coupling to the SM Higgs scales as,
\be
\lambda_h^{(1)} \propto ( 1 + \sin 2 \beta  \sin 2 \theta_{_{\rm TB}}) m_{\x_1} - ( \sin 2 \beta + \sin 2 \theta_{_{\rm TB}} ) M_B
,
\ee
leading to a direct detection blind spot under the condition:
\be
\sin 2 \theta_{_{\rm TB}} \approx \frac{m_{\x_1} - M_B \sin 2 \beta} {M_B - m_{\x_1} \sin 2 \beta}
~.
\ee
In the limit that $M_B \gg M_T \sim m_{\x_1}$~, $\lambda_h^{(1)} \propto s_{2 \beta} + s_{2 \theta_{_{\rm TB}}}$ and no blind spot exists. In contrast, cancellations are possible when $M_T \gg M_B \sim m_{\x_1}$~, which implies $\lambda_h^{(1)} \propto (1-s_{2 \beta}) (1-s_{2 \theta_{_{\rm TB}}})$. The latter case explains the lack of sensitivity for LZ when $M_T \gg M_B$ and $\tan\theta_{_{\rm TB}}>0$ in Fig.~\ref{fig:TripBid1}. 

Any viable dark matter candidate must be electrically neutral. Throughout the parameter space of Fig.~\ref{fig:TripBid1}, however, the lightest neutral and charged states are nearly degenerate at tree-level, implying that radiative corrections are potentially relevant. As described in Appendix~\ref{sec:loops}, we calculate the one-loop corrections to the dark sector masses and investigate whether this leads to $m_{\x_1^\pm} < m_{\x_1}$. Setting the $\overline{\text{MS}}$ renormalization scale to $\mu = 1$ TeV, we find that $m_{\x_1^\pm} > m_{\x_1}$ in the regions shown in Fig.~\ref{fig:TripBid1}. Alternatively, for larger degrees of decoupling, $M_T \ll M_B \gtrsim \mathcal{O}(10)$ TeV or $M_B \ll M_T \gtrsim \mathcal{O}(10)$ TeV, radiative corrections lead to $m_{\x_1^\pm} < m_{\x_1}$ when the remaining parameters are set to the values given in Fig.~\ref{fig:TripBid1}.

\subsection{Indirect Detection}

Constraints from searches for the annihilation products of dark matter are not particularly stringent in the class of models presented here. Gamma-ray observations of dwarf galaxies~\cite{Ackermann:2015zua} and the Galactic Center~\cite{Hooper:2012sr} are currently only sensitive to thermal relics with masses below $\sim$100 GeV. Although measurements of the cosmic ray anti-proton spectrum can provide a competitive constraint over a similar mass range~\cite{Giesen:2015ufa,Jin:2015sqa,Hooper:2014ysa}, interpretations of cosmic ray data currently involve significant astrophysical uncertainties.  

In some of the parameter space considered in this paper, the low-velocity dark matter annihilation cross section may experience non-negligible Sommerfeld enhancements, most notably boosting the annihilation rate to distinctive $\gamma \gamma$ and $\gamma Z$ final states. More specifically, if the dark matter is largely bidoublet-like, Sommerfeld enhancements can result from the couplings to the $W^{\pm}$, similar to the case of a Higgsino-like neutralino. Even with this enhancement, however, the predicted gamma-ray signal remains beyond the reach of current or next generation telescopes. Note that in none of the models discussed here does the dark matter experience a large Sommerfeld enhancement of the type predicted for a wino-like neutralino~\cite{Cohen:2013ama,Fan:2013faa} (as none of the models include a $SU(2)_L$ triplet).

\subsection{Dark Matter Stability}
\label{sec:stability}

Throughout this paper, we have implicitly defined Lagrangians above the scale $v_R \sim 3-4$ TeV, which breaks $SU(2)_L \times SU(2)_R \times U(1)_{B-L}$ down to $SU(2)_L \times U(1)_Y$. Furthermore, the examples in Secs.~\ref{sec:singlettriplet}-\ref{sec:tripletbidoublet} all possess an accidental parity symmetry, under which the fermions of the dark sector are odd. This can be understood from the fact that $v_R$ breaks $U(1)_{B-L}$ down to a non-trivial $\mathbb{Z}_2$ subgroup, and as a result, the lightest new fermion with even $B-L$ is automatically stable~\cite{Heeck:2015qra}. 

However, above the scale $v_R$, this stabilizing symmetry need not be respected. In particular, new physics not contained in a minimal left-right symmetric model may generate interactions that allow for the lightest parity-odd fermion to decay. In this section, we outline a simple argument demonstrating that any new physics that respects $B-L$ and Lorentz invariance will not generate such processes as long as the $B-L$ charge of the dark matter multiplet is chosen appropriately.

We begin by assuming that the multiplet, $X$, is some fermionic dark matter gauge eigenstate, similar to the ones described in Sec.~\ref{sec:mixedmodels}, with $B-L$ charge $Q$. The neutral component of $X$ ($\x$) is assumed to be the cosmological dark matter. Gauge and Lorentz invariance dictate that $\x$ may decay only through an operator of the form
\be
O = X \times (\text{any \# of bosons}) \times (\text{odd \# of SM fermions})
~.
\ee
Next, we let the quantity ``$(\text{odd \# of SM fermions}) $" consist of $n_l$ SM lepton fields (each with $B-L=\pm 1$) and $n_q$ SM quarks (each with $B-L=\pm 1/3$), and imagine that some subset of the lepton and/or quark fields cancel in $B-L$. We will denote the number of uncanceled leptons/quarks by $n_{l,q}^\prime$. This cancellation can only take place among an even number of fields, and hence there are still an odd number of uncanceled SM fermions in the product. Furthermore, at or below the scale $v_R$, all the bosons of a left-right model are evenly charged under $B-L$. These two insights imply the following system of equations:
\begin{align}
n_l^\prime+n_q^\prime &= 2 n+1 \quad (\text{odd \# of SM fermions})
\nl
Q+2m&+n_l^\prime + \frac{1}{3} n_q^\prime = 0 \quad (\text{$B-L$ invariant})
,
\end{align}
where $m$ and $n$ are some integers. Solving for $n_{l,q}^\prime$ yields:
\begin{align}
n_l^\prime &= \frac{-1}{2} (1+3Q+6m+2n)
\nl
n_q^\prime &= \frac{3}{2} (1+2(m+n)+Q)
~.
\end{align}
 $n_{l,q}^\prime$ are integers by definition, and hence the second line above implies that $\x$ may only decay if $3 Q$ is an odd integer. Therefore, we have shown that the lightest neutral component, $\x$, of a dark matter multiplet, $X$, is exactly stable, if $X$ does \emph{not} possess a $B-L$ charge of $Q = \pm (1,3,5,\ldots)/3$.


\section{Summary and Conclusions}
\label{summary}

Motivated by the Run 1 diboson and dijet excesses at ATLAS and CMS, we have explored dark matter in left-right symmetric models, including those in which the dark matter candidate is a mixed state of fermionic multiplets. Such models are limited to singlet-triplet, singlet-bidoublet, and triplet-bidoublet dark matter, and we find that an acceptable thermal relic abundance can be obtained for a wide range of masses in each of these cases. New gauge and Higgs bosons present in minimal left-right symmetric models provide the dominant interactions between the SM and dark sector, while scattering in direct detection experiments is largely governed by the tree-level exchange of a SM Higgs or $Z^\prime$. Interestingly, stability of the lightest neutral state in the dark sector is guaranteed to all orders by $B-L$ gauge invariance for the models considered. 

The parameter space of the singlet-triplet model will be largely unconstrained by future experiments such as LUX-ZEPLIN and XENON1T. In contrast, these experiments will be able to significantly investigate models of thermal dark matter in the singlet-bidoublet and triplet-bidoublet cases. We have also taken note of interesting regions of parameter space where the branching fraction $\text{BR}(W^\prime \to \text{dark sector}) \gtrsim 10 \%$. Such large invisible branching fractions help move us in the direction of being able to fit the observed Run 1 diboson signal at values of $g_R$ closer to the theoretically attractive choice of $g_R = g_L$.

As well as investigating the phenomenology of dark matter in left-right models, we have also supplied a comprehensive set of Appendices that contain all the information necessary to investigate these, and related models, further in the future.

\bigskip
\bigskip

\textbf{Acknowledgments.} We would like to thank Anthony DiFranzo for valuable discussions. AB is supported by the Kavli Institute for Cosmological Physics at the University of Chicago through grant NSF PHY-1125897. DH is supported by the US Department of Energy under contract DE-FG02-13ER41958. GM is supported by the Fermilab Graduate Student Research Program in Theoretical Physics and in part by the National Research Foundation of South Africa, Grant No. 88614. Fermilab is operated by Fermi Research Alliance, LLC, under Contract No. DE-AC02-07CH11359 with the US Department of Energy. 

\appendix

\section{Gauge Boson Couplings}
\label{appA}

In the $v_R \gg v$ limit, the couplings of the $Z^\prime$ and $W^\prime$ to SM fermions are as follows:
\begin{align}
\mathcal{L}_{V^\prime f} &= \sum\limits_f Z^\prime_\mu ~ \bar{f} \gamma^\mu ( g_f^{\prime v} + g_f^{\prime a} \gamma^5 ) f + \frac{g_R}{2 \sqrt{2}}~ W_\mu^{\prime +} ~ \bar{u} \gamma^\mu (1+\gamma^5) d +  \frac{g_R}{2 \sqrt{2}}~ W_\mu^{\prime -} ~ \bar{d} \gamma^\mu (1+\gamma^5) u 
\nl
&+ \frac{g_R}{2 \sqrt{2}}~ W_\mu^{\prime +} ~ \bar{\nu} \gamma^\mu (1+\gamma^5) l +  \frac{g_R}{2 \sqrt{2}}~ W_\mu^{\prime -} ~ \bar{l} \gamma^\mu (1+\gamma^5) \nu,
\end{align}
where
\begin{align}
g_u^{\prime v} &\equiv \frac{3 g_R^2-5 g^{\prime 2}}{12 \sqrt{g_R^2-g^{\prime 2}}} ~,~~ g_u^{\prime a} \equiv \frac{1}{4} \sqrt{g_R^2-g^{\prime 2}} ~,~~ g_d^{\prime v} \equiv \frac{g^{\prime 2} - 3 g_R^2}{12 \sqrt{g_R^2-g^{\prime 2}}} ~,~~ g_d^{\prime a} \equiv \frac{-1}{4} \sqrt{g_R^2-g^{\prime 2}}
\nl
g_\nu^{\prime v} &\equiv \frac{g^{\prime 2}}{4 \sqrt{g_R^2-g^{\prime 2}}} ~,~~ g_\nu^{\prime a} \equiv \frac{-g^{\prime 2}}{4 \sqrt{g_R^2-g^{\prime 2}}} ~,~~ g_l^{\prime v} \equiv \frac{3 g^{\prime 2} - g_R^2}{4 \sqrt{g_R^2 - g^{\prime 2}}} ~,~~ g_l^{\prime a} \equiv \frac{-1}{4} \sqrt{g_R^2-g^{\prime 2}}
.
\end{align}

The non-vanishing cubic self-interaction terms involving non-SM gauge bosons are given as follows:
\begin{eqnarray}
\mathcal{L}_{W^\prime W^\prime \gamma} &=& -i g_R c_w s_R \left( ~ \partial_\nu A_\mu ~ W^{\prime - \nu} ~ W^{\prime + \mu} + \partial_\nu W_\mu^{\prime -} ~ A^{[\mu} W^{\prime + \nu]} ~ \right) + \text{h.c.}, \\
\mathcal{L}_{W^\prime W^\prime Z} &=& i g_R s_w s_R \left( Z_\mu ~ W_\nu^{\prime +} ~ \partial^{[\nu} W^{\prime - \mu]} + \partial_\nu Z_\mu ~ W^{\prime - \nu} ~ W^{\prime + \mu} \right) + \text{h.c.},  
\nl
\mathcal{L}_{W^\prime W^\prime Z^\prime} &=& -i g_R c_R \left( Z^\prime_\mu ~ W_\nu^{\prime +} ~ \partial^{[ \nu}W^{\prime - \mu]} + \partial_\nu Z^\prime_\mu ~ W^{\prime - \nu} ~ W^{\prime + \mu} \right) + \text{h.c.},
\nl 
\mathcal{L}_{W^\prime W Z} &=& \frac{i g_L}{c_W} \sin{\theta_+} \Big[  Z_\mu \left( W_\nu^{\prime +} \partial^{[ \nu} W^{- \mu ]} + W_\nu^+ \partial^{[ \nu} W^{\prime - \mu ]} \right) + \partial_\nu Z_\mu W^{\prime + [ \mu} W^{- \nu ]}  \Big] + \text{h.c.} ~. \nonumber
\end{eqnarray}

And lastly, we list the non-SM cubic gauge-Higgs interactions terms, to leading order in $m_W/m_{W^\prime}$:
\begin{align}
\mathcal{L}_{Z \phi_1 \phi_2} &= \frac{-g_L}{2 c_w} ~ Z^\mu ~ \left( H ~ \partial_\mu A - A ~ \partial_\mu H \right) - \frac{i}{2} \left( g_L c_w - g^\prime s_w \right) Z^\mu \left( H^+ ~ \partial_\mu H^- - H^- ~ \partial_\mu H^+ \right), 
\nl
\mathcal{L}_{Z^\prime \phi_1 \phi_2} &= \frac{g_R c_R}{2} ~ Z^{\prime \mu} ~ \left( H ~ \partial_\mu A - A ~ \partial_\mu H \right) 
- \frac{i g_R c_R}{2} Z^{\prime \mu} \left( H^+ ~ \partial_\mu H^- - H^- ~ \partial_\mu H^+ \right), 
\nl
\mathcal{L}_{W \phi_1 \phi_2} &= 
\frac{i g_L}{2} ~ W^{+ \mu} \left( H ~ \partial_\mu H^- - H^- ~ \partial_\mu H \right) + \frac{g_L}{2} ~ W^{+ \mu} ~ \left( A ~ \partial_\mu H^- - H^- ~ \partial_\mu A \right) + \text{h.c.},
\nl
\mathcal{L}_{W^\prime \phi_1 \phi_2} &= \frac{i g_R c_{2 \beta}}{2} ~ W^{\prime + \mu} ~ \left( h ~ \partial_\mu H^- - H^- ~ \partial_\mu h \right) + \frac{i g_R s_{2 \beta}}{2} ~ W^{\prime + \mu} ~ \left( H ~ \partial_\mu H^- - H^- ~ \partial_\mu H \right)
\nl
&+ \frac{g_R s_{2 \beta}}{2} ~ W^{\prime + \mu} \left( H^- ~ \partial_\mu A - A ~ \partial_\mu H^- \right) + \text{h.c.},
\nl
\mathcal{L}_{\gamma \phi_1 \phi_2} &= i e ~ A^\mu ~ \left( H^- ~ \partial_\mu H^+ - H^+ ~ \partial_\mu H^- \right), \nonumber \end{align}
\begin{align}
\mathcal{L}_{V V \phi} &= \bigg[ g_R m_W ~ W^{\prime +}_\mu W^{- \mu} ~ \left( - s_{2 \beta}~h + c_{2 \beta}~H + i c_{2 \beta}~A  \right) - \frac{g_R (g_R^2 + g^{\prime 2})}{g_L \sqrt{g_R^2 - g^{\prime 2}}} ~ m_W ~ W^{\prime + \mu} Z_\mu^\prime (c_{2 \beta} H^- )
\nl
&- \frac{g_R m_W c_{2 \beta}}{c_w} ~ W^{\prime + \mu} H^-  Z_\mu  + \text{h.c.} \bigg] + \frac{g_R^2}{g_L} m_W ~ h ~ W^{\prime + \mu} ~ W_\mu^{\prime -}  - \frac{g_R c_R m_W}{c_W} ~ h ~ Z^{\prime \mu} Z_\mu \nonumber \\
&+ \frac{g_R^2 c_R^2 m_W}{2 g_L} ~ h ~ Z^{\prime \mu} Z^\prime_\mu. \end{align}

\section{Singlet-Triplet Dark Matter Interactions with Gauge or Higgs Bosons}
\label{appB}

The leading order interactions of gauge bosons with the dark sector are written as:
\begin{align}
\mathcal{L}_{\x V} &= \sum\limits_i ~ g_{Z^\prime}^{(i)} ~ Z_\mu^\prime ~ \xiibar \gamma^\mu \gamma^5 \xii + \sum\limits_{i<j} Z_\mu^\prime ~ \xiibar \gamma^\mu \left( i g_{Z^\prime v}^{(ij)} + g_{Z^\prime a}^{(ij)} \gamma^5 \right) \xjj
\nl
&+  \sum\limits_i ~ W_\mu^+ ~ \xpbar \gamma^\mu \left( g_{W v}^{(i)} + g_{W a}^{(i)} \gamma^5 \right) \xii + W_\mu^- ~ \xiibar \gamma^\mu \left( g_{W v}^{(i)*} + g_{W a}^{(i)*} \gamma^5 \right) \xp
\nl
&+  \sum\limits_i ~ W_\mu^{\prime +} ~ \xpbar \gamma^\mu \left( g_{W^\prime v}^{(i)} + g_{W^\prime a}^{(i)} \gamma^5 \right) \xii + W_\mu^{\prime -} ~ \xiibar \gamma^\mu \left( g_{W^\prime v}^{(i)*} + g_{W^\prime a}^{(i)*} \gamma^5 \right) \xp
\nl
&+ e ~ A_\mu ~ \xpbar \gamma^\mu \xp + g_Z^{(+)} ~ Z_\mu ~ \xpbar \gamma^\mu \xp + g_{Z^\prime}^{(+)} ~ Z_\mu^\prime ~ \xpbar \gamma^\mu \xp
\nl
& -g_R ~ W_\mu^{\prime +} ~ \xppbar \gamma^\mu \xp - g_R ~ W_\mu^{\prime -} ~ \xpbar \gamma^\mu \xpp
\nl
& -g_R \sin{\theta_+}~ W_\mu^+ ~ \xppbar \gamma^\mu \xp - g_R \sin{\theta_+}~ W_\mu^- ~ \xpbar \gamma^\mu \xpp
\nl
& + 2 e ~ A_\mu ~ \xppbar \gamma^\mu \xpp + g_Z^{(++)} ~ Z_\mu ~ \xppbar \gamma^\mu \xpp + g_{Z^\prime}^{(++)} ~ Z_\mu^\prime ~ \xppbar \gamma^\mu \xpp,
\end{align}
where we have defined the couplings as follows:
\begin{align}
g_{Z^\prime}^{(i)} &\equiv \frac{g_R}{2 c_R} \left( | N_{t_1}^i |^2 - | N_{t_2}^i |^2 \right),
\nl
g_{Z^\prime v}^{(ij)} &\equiv \frac{g_R}{c_R} ~ \text{Im}\left( N_{t_1}^i N_{t_1}^{j *} - N_{t_2}^i N_{t_2}^{j *} \right),
\nl
g_{Z^\prime a}^{(ij)} &\equiv \frac{g_R}{c_R} ~ \text{Re}\left( N_{t_1}^i N_{t_1}^{j *} - N_{t_2}^i N_{t_2}^{j *} \right),
\nl
g_{W v}^{(i)} &\equiv \frac{g_R \sin{\theta_+}}{2} \left( N_{t_1}^{i} + N_{t_2}^{i*} \right),
\nl
 g_{W a}^{i} &\equiv \frac{-g_R \sin{\theta_+}}{2} \left( N_{t_1}^{i} - N_{t_2}^{i*} \right),
\nl
g_{W^\prime v}^{(i)} &\equiv \frac{g_R}{2} \left( N_{t_1}^{i} + N_{t_2}^{i*} \right),
\nl
g_{W^\prime a}^{(i)} &\equiv \frac{-g_R}{2} \left( N_{t_1}^{i} - N_{t_2}^{i*} \right),
\nl
g_Z^{(+)} &\equiv -g^\prime s_w,
\nl
 g_{Z^\prime}^{(+)} &\equiv \frac{-g^{\prime 2}}{\sqrt{g_R^2 - g^{\prime 2}}},
\nl
g_Z^{(++)} &\equiv - 2 g^\prime s_w,
\nl
 g_{Z^\prime}^{(++)} &\equiv \frac{g_R^2 - 2 g^{\prime 2}}{\sqrt{g_R^2 - g^{\prime 2}}}.
\end{align}

Similarly, the interactions with the Higgs sector are given by:
\begin{align}
\mathcal{L}_{\x \phi} &= \sum\limits_i ~ \Delta^{++} ~ \xppbar \left( \lambda_{++s}^{(i)} + \lambda_{++p}^{(i)} ~ \gamma^5 \right) \xii + \Delta^{--} ~ \xiibar \left( \lambda_{++s}^{(i)*} - \lambda_{++p}^{(i)*} ~ \gamma^5 \right) \xpp
\nl
&+ \sum\limits_i \lambda_0^{(i)} ~ \Delta^0 ~ \xiibar \xii + \sum\limits_{i < j} \Delta^0 ~ \xiibar \left(  \lambda_{0s}^{(ij)} + \lambda_{0p}^{(ij)} i \gamma^5 \right) \xjj
,
\end{align}
where
\begin{align}
\lambda_{++s}^{(i)} &\equiv \frac{-1}{2} \left( \lambda_1 N_S^{i *} + \lambda_2 N_S^i \right),
\nl
\lambda_{++p}^{(i)} &\equiv \frac{-1}{2} \left( \lambda_1 N_S^{i *} - \lambda_2 N_S^i \right),
\nl
\lambda_0^{(i)} &\equiv \frac{-1}{\sqrt{2}} ~ \text{Re} \Big[ N_S^i \left( \lambda_1 N_{t_1}^i + \lambda_2 N_{t_2}^i \right) \Big],
\nl
\lambda_{0s}^{(ij)} &\equiv \frac{-1}{\sqrt{2}} ~ \text{Re} \Big[ \lambda_1 \left( N_S^i N_{t_1}^j + N_S^j N_{t_1}^i \right) + \lambda_2 \left( N_S^i N_{t_2}^j + N_S^j N_{t_2}^i \right) \Big],
\nl
\lambda_{0p}^{(ij)} &\equiv \frac{1}{\sqrt{2}} ~ \text{Im} \Big[ \lambda_1 \left( N_S^i N_{t_1}^j + N_S^j N_{t_1}^i \right) + \lambda_2 \left( N_S^i N_{t_2}^j + N_S^j N_{t_2}^i \right) \Big].
\end{align}
Note that there are no charged fermion-charged fermion-Higgs interactions, because singlet mixing is needed in order to induce a coupling to the Higgs sector. In practice, we will assume that the physical triplet-Higgs content is decoupled.

\section{Singlet-Bidoublet Dark Matter Interactions with Gauge or Higgs Bosons}
\label{appC}

The leading order interactions of gauge bosons with the dark sector are written as:
\begin{align}
\mathcal{L}_{\x V} &= \sum\limits_i ~ \left( g_Z^{(i)} ~ Z_\mu + g_{Z^\prime}^{(i)} ~ Z_\mu^\prime \right)~ \xiibar \gamma^\mu \gamma^5 \xii 
\nl
&+  \sum\limits_{i<j} Z_\mu ~ \xiibar \gamma^\mu \left( i g_{Z v}^{(ij)} + g_{Z a}^{(ij)} \gamma^5 \right) \xjj + \sum\limits_{i<j} Z_\mu^\prime ~ \xiibar \gamma^\mu \left( i g_{Z^\prime v}^{(ij)} + g_{Z^\prime a}^{(ij)} \gamma^5 \right) \xjj
\nl
&+  \sum\limits_i ~ W_\mu^{+} ~ \xpbar \gamma^\mu \left( g_{W v}^{(i)} + g_{W a}^{(i)} \gamma^5 \right) \xii + W_\mu^{-} ~ \xiibar \gamma^\mu \left( g_{W v}^{(i)*} + g_{W a}^{(i)*} \gamma^5 \right) \xp
\nl
&+  \sum\limits_i ~ W_\mu^{\prime +} ~ \xpbar \gamma^\mu \left( g_{W^\prime v}^{(i)} + g_{W^\prime a}^{(i)} \gamma^5 \right) \xii + W_\mu^{\prime -} ~ \xiibar \gamma^\mu \left( g_{W^\prime v}^{(i)*} + g_{W^\prime a}^{(i)*} \gamma^5 \right) \xp
\nl
&+ e ~ A_\mu ~ \xpbar \gamma^\mu \xp + g_Z^{(+)} ~ Z_\mu ~ \xpbar \gamma^\mu \xp + g_{Z^\prime}^{(+)} ~ Z_\mu^\prime ~ \xpbar \gamma^\mu \xp,
\end{align}
where we have defined the couplings as follows:
\begin{align}
g_Z^{(i)} &\equiv \frac{-g_L}{4 c_w} \left( | N_{b_1}^i |^2 - | N_{b_2}^i |^2 \right), 
\nl
g_{Z^\prime}^{(i)} &\equiv \frac{1}{4} ~ \sqrt{g_R^2-g^{\prime 2}} ~ \left( | N_{b_1}^i |^2 - | N_{b_2}^i |^2 \right),
\nl
g_{Z v}^{(ij)} &\equiv \frac{-g_L}{2 c_w} ~ \text{Im}\left( N_{b_1}^i N_{b_1}^{j *} - N_{b_2}^i N_{b_2}^{j *}\right),
\nl
g_{Z a}^{(ij)} &\equiv \frac{-g_L}{2 c_w} ~ \text{Re}\left( N_{b_1}^i N_{b_1}^{j *} - N_{b_2}^i N_{b_2}^{j *}\right),
\nl
g_{Z^\prime v}^{(ij)} &\equiv \frac{1}{2} \sqrt{g_R^2-g^{\prime 2}} ~ \text{Im}\left( N_{b_1}^i N_{b_1}^{j *} - N_{b_2}^i N_{b_2}^{j *}\right),
\nl
g_{Z^\prime a}^{(ij)} &\equiv \frac{1}{2} \sqrt{g_R^2-g^{\prime 2}} ~ \text{Re}\left( N_{b_1}^i N_{b_1}^{j *} - N_{b_2}^i N_{b_2}^{j *}\right),
\nl
g_{W v}^{(i)} &\equiv \frac{-g_L}{2\sqrt{2}} \left( N_{b_1}^{i*} + N_{b_2}^{i} \right) ,
\nl
g_{W a}^{(i)} &\equiv \frac{-g_L}{2\sqrt{2}} \left( N_{b_1}^{i*} - N_{b_2}^{i} \right),
\nl
g_{W^\prime v}^{(i)} &\equiv \frac{g_R}{2\sqrt{2}} \left( N_{b_1}^{i} + N_{b_2}^{i*} \right) ,
\nl
 g_{W^\prime a}^{(i)} &\equiv \frac{-g_R}{2\sqrt{2}} \left( N_{b_1}^{i} - N_{b_2}^{i*} \right),
\nl
g_Z^{(+)} &\equiv \frac{1}{2} \left( g_L c_w - g^\prime s_w \right) ,
\nl
g_{Z^\prime}^{(+)} &\equiv \frac{1}{2} ~ \sqrt{g_R^2 - g^{\prime 2}}.
\end{align}

Similarly, the dark sector interacts with the Higgs bosons of the bidoublet as follows:
\begin{align}
\mathcal{L}_{\x \phi} &=  \sum\limits_i ~ \left( \lambda_h^{(i)}  ~ h + \lambda_H^{(i)}  ~ H \right) ~ \xiibar \xii +  \lambda_A^{(i)}   A  ~ \xiibar i \gamma^5 \xii
\nl
&+ \sum\limits_{i < j} ~ h ~ \xiibar \left(  \lambda_{hs}^{(ij)} + \lambda_{hp}^{(ij)} ~ i \gamma^5 \right) \xjj + H ~ \xiibar \left(  \lambda_{Hs}^{(ij)} + \lambda_{Hp}^{(ij)} ~ i \gamma^5 \right) \xjj + A ~ \xiibar \left(  \lambda_{As}^{(ij)} + \lambda_{Ap}^{(ij)} ~ i \gamma^5 \right) \xjj 
\nl
&+ \sum\limits_i ~ H^+ ~ \xpbar \left( \lambda_{H^+s}^{(i)} + \lambda_{H^+p}^{(i)} ~ \gamma^5 \right) \xii + H^- ~ \xiibar \left( \lambda_{H^+s}^{(i)*} - \lambda_{H^+p}^{(i)*} ~ \gamma^5 \right) \xp,
\end{align}
where 
\begin{align}
\lambda_h^{(i)} &= \frac{-\lambda}{\sqrt{2}} ~ \text{Re}\Big[ N_s^i (c_\beta N_{b_1}^i + s_\beta N_{b_2}^i )\Big] -  \frac{\tilde{\lambda}}{\sqrt{2}} ~ \text{Re}\Big[ N_s^i (s_\beta N_{b_1}^i + c_\beta N_{b_2}^i )\Big],
\nl
\lambda_H^{(i)} &= \frac{-\lambda}{\sqrt{2}} ~ \text{Re}\Big[ N_s^i (s_\beta N_{b_1}^i - c_\beta N_{b_2}^i )\Big] +  \frac{\tilde{\lambda}}{\sqrt{2}} ~ \text{Re}\Big[ N_s^i ( c_\beta N_{b_1}^i - s_\beta N_{b_2}^i )\Big],
\nl
\lambda_A^{(i)} &= \frac{-\lambda}{\sqrt{2}} ~ \text{Re}\Big[ N_s^i (s_\beta N_{b_1}^i + c_\beta N_{b_2}^i )\Big] +  \frac{\tilde{\lambda}}{\sqrt{2}} ~ \text{Re}\Big[ N_s^i (c_\beta N_{b_1}^i + s_\beta N_{b_2}^i )\Big],
\nl
\lambda_{hs}^{(ij)} &= \frac{-\lambda}{\sqrt{2}} ~ \text{Re}\Big[ c_\beta \left( N_S^i N_{b_1}^j + N_S^j N_{b_1}^i \right) + s_\beta \left( N_S^i N_{b_2}^j + N_S^j N_{b_2}^i \right) \Big]
\nl
&~~~ - \frac{\tilde{\lambda}}{\sqrt{2}} ~ \text{Re}\Big[ s_\beta \left( N_S^i N_{b_1}^j + N_S^j N_{b_1}^i \right) + c_\beta \left( N_S^i N_{b_2}^j + N_S^j N_{b_2}^i \right) \Big],
\nl
\lambda_{hp}^{(ij)} &= \frac{\lambda}{\sqrt{2}} ~ \text{Im}\Big[ c_\beta \left( N_S^i N_{b_1}^j + N_S^j N_{b_1}^i \right) + s_\beta \left( N_S^i N_{b_2}^j + N_S^j N_{b_2}^i \right) \Big]
\nl
&~~~ + \frac{\tilde{\lambda}}{\sqrt{2}} ~ \text{Im}\Big[ s_\beta \left( N_S^i N_{b_1}^j + N_S^j N_{b_1}^i \right) + c_\beta \left( N_S^i N_{b_2}^j + N_S^j N_{b_2}^i \right) \Big], 
\nl
\lambda_{Hs}^{(ij)} &= \frac{-\lambda}{\sqrt{2}} ~ \text{Re}\Big[ s_\beta \left( N_S^i N_{b_1}^j + N_S^j N_{b_1}^i \right) - c_\beta \left( N_S^i N_{b_2}^j + N_S^j N_{b_2}^i \right) \Big] 
\nl
&~~~+ \frac{\tilde{\lambda}}{\sqrt{2}} ~ \text{Re}\Big[ c_\beta \left( N_S^i N_{b_1}^j + N_S^j N_{b_1}^i \right) - s_\beta \left( N_S^i N_{b_2}^j + N_S^j N_{b_2}^i \right) \Big],
\nl
\lambda_{Hp}^{(ij)} &= \frac{\lambda}{\sqrt{2}} ~ \text{Im}\Big[ s_\beta \left( N_S^i N_{b_1}^j + N_S^j N_{b_1}^i \right) - c_\beta \left( N_S^i N_{b_2}^j + N_S^j N_{b_2}^i \right) \Big] 
\nl
&~~~- \frac{\tilde{\lambda}}{\sqrt{2}} ~ \text{Im}\Big[ c_\beta \left( N_S^i N_{b_1}^j + N_S^j N_{b_1}^i \right) - s_\beta \left( N_S^i N_{b_2}^j + N_S^j N_{b_2}^i \right) \Big],
\nl
\lambda_{As}^{(ij)} &= \frac{-\lambda}{\sqrt{2}} ~ \text{Im}\Big[ s_\beta \left( N_S^i N_{b_1}^j + N_S^j N_{b_1}^i \right) + c_\beta \left( N_S^i N_{b_2}^j + N_S^j N_{b_2}^i \right) \Big] 
\nl
&~~~+ \frac{\tilde{\lambda}}{\sqrt{2}} ~ \text{Im}\Big[ c_\beta \left( N_S^i N_{b_1}^j + N_S^j N_{b_1}^i \right) + s_\beta \left( N_S^i N_{b_2}^j + N_S^j N_{b_2}^i \right) \Big],
\nl
\lambda_{Ap}^{(ij)} &= \frac{-\lambda}{\sqrt{2}} ~ \text{Re}\Big[ s_\beta \left( N_S^i N_{b_1}^j + N_S^j N_{b_1}^i \right) + c_\beta \left( N_S^i N_{b_2}^j + N_S^j N_{b_2}^i \right) \Big] 
\nl
&~~~+ \frac{\tilde{\lambda}}{\sqrt{2}} ~ \text{Re}\Big[ c_\beta \left( N_S^i N_{b_1}^j + N_S^j N_{b_1}^i \right) + s_\beta \left( N_S^i N_{b_2}^j + N_S^j N_{b_2}^i \right) \Big]
\nl
&~~~+ \frac{\tilde{\lambda}}{\sqrt{2}} ~ \text{Im}\Big[ s_\beta \left( N_S^i N_{b_1}^j + N_S^j N_{b_1}^i \right) - c_\beta \left( N_S^i N_{b_2}^j + N_S^j N_{b_2}^i \right) \Big]
\nl
&~~~+ \frac{\tilde{\lambda}}{\sqrt{2}} ~ \text{Re}\Big[ s_\beta \left( N_S^i N_{b_1}^j + N_S^j N_{b_1}^i \right) - c_\beta \left( N_S^i N_{b_2}^j + N_S^j N_{b_2}^i \right) \Big],
\nl
\lambda_{H^+s}^{(i)} &= \frac{- \lambda}{2} \left( s_\beta N_S^i - c_\beta N_S^{i *} \right) + \frac{\tilde{\lambda}}{2} \left( c_\beta N_S^i - s_\beta N_S^{i *} \right),
\nl
\lambda_{H^+p}^{(i)} &= \frac{\lambda}{2} \left( s_\beta N_S^i + c_\beta N_S^{i *} \right) - \frac{\tilde{\lambda}}{2} \left( c_\beta N_S^i + s_\beta N_S^{i *} \right).
\end{align}
Note that there are no charged fermion-charged fermion-Higgs interactions, as singlet mixing is required to induce a coupling to the Higgs sector.

\section{Triplet-Bidoublet Dark Matter Interactions with Gauge or Higgs Bosons}
\label{appD}

The leading order interactions of gauge bosons with the dark sector are written in this model as:
\begin{align}
\mathcal{L}_{\x V} &= \sum\limits_i ~ \left( g_Z^{(i)} ~ Z_\mu + g_{Z^\prime}^{(i)} ~ Z_\mu^\prime \right)~ \xiibar \gamma^\mu \gamma^5 \xii 
\nl
&+  \sum\limits_{i<j} Z_\mu ~ \xiibar \gamma^\mu \left( i g_{Z v}^{(ij)} + g_{Z a}^{(ij)} \gamma^5 \right) \xjj + \sum\limits_{i<j} Z_\mu^\prime ~ \xiibar \gamma^\mu \left( i g_{Z^\prime v}^{(ij)} + g_{Z^\prime a}^{(ij)} \gamma^5 \right) \xjj
\nl
&+  \sum\limits_{i,\, j} ~ W_\mu^{+} ~ \overline{\xp_i} \gamma^\mu \left( g_{W v}^{(ij)} + g_{W a}^{(ij)} \gamma^5 \right) \xjj + W_\mu^{-} ~ \xjjbar \gamma^\mu \left( g_{W v}^{(ij)*} + g_{W a}^{(ij)*} \gamma^5 \right) \xp_i
\nl
&+  \sum\limits_{i,\, j} ~ W_\mu^{\prime +} ~ \overline{\xp_i} \gamma^\mu \left( g_{W^\prime v}^{(ij)} + g_{W^\prime a}^{(ij)} \gamma^5 \right) \xjj + W_\mu^{\prime -} ~ \xjjbar \gamma^\mu \left( g_{W^\prime v}^{(ij)*} + g_{W^\prime a}^{(ij)*} \gamma^5 \right) \xp_i
\nl
&+ \sum\limits_i ~ e ~ A_\mu ~ \overline{\xp_i} \gamma^\mu \xp_i + Z_\mu ~ \overline{\xp_i} \gamma^\mu \left(  g_{Zv}^{(i +)} + g_{Za}^{(i +)} \gamma^5  \right) \xp_i +  Z_\mu^\prime ~ \overline{\xp_i} \gamma^\mu \left(  g_{Z^\prime v}^{(i +)} + g_{Z^\prime a}^{(i +)} \gamma^5  \right) \xp_i
\nl
&+ \bigg[ \sum\limits_{i < j} ~ Z_\mu ~ \overline{\xp_i} \gamma^\mu \left( g_{Zv}^{(i j +)} + g_{Za}^{(i j +)} \gamma^5\right) \xp_j +  Z_\mu^\prime ~ \overline{\xp_i} \gamma^\mu \left( g_{Z^\prime v}^{(i j +)} + g_{Z^\prime a}^{(i j +)} \gamma^5\right) \xp_j + \text{h.c.} \bigg]
,
\end{align}
where we have defined the couplings as follows:
\begin{align}
g_Z^{(i)} &\equiv \frac{-g_L}{4 c_w} \left( | N_{b_1}^i |^2 - | N_{b_2}^i |^2 \right),
\nl
g_{Z^\prime}^{(i)} &\equiv \frac{1}{4} ~ \sqrt{g_R^2-g^{\prime 2}} ~ \left( | N_{b_1}^i |^2 - | N_{b_2}^i |^2 \right),
\nl
g_{Z v}^{(ij)} &\equiv \frac{-g_L}{2 c_w} ~ \text{Im}\left( N_{b_1}^i N_{b_1}^{j *} - N_{b_2}^i N_{b_2}^{j *}\right),
\nl g_{Z a}^{(ij)} &\equiv \frac{-g_L}{2 c_w} ~ \text{Re}\left( N_{b_1}^i N_{b_1}^{j *} - N_{b_2}^i N_{b_2}^{j *}\right),
\nl
g_{Z^\prime v}^{(ij)} &\equiv \frac{1}{2} \sqrt{g_R^2-g^{\prime 2}} ~ \text{Im}\left( N_{b_1}^i N_{b_1}^{j *} - N_{b_2}^i N_{b_2}^{j *}\right),
\nl
g_{Z^\prime a}^{(ij)} &\equiv \frac{1}{2} \sqrt{g_R^2-g^{\prime 2}} ~ \text{Re}\left( N_{b_1}^i N_{b_1}^{j *} - N_{b_2}^i N_{b_2}^{j *}\right),
\nl
g_{W v}^{(ij)} &\equiv \frac{-g_L}{2 \sqrt{2}} ~ \left( U_{2i} ~ N_{b_1}^{j*} + V_{2i} ~ N_{b_2}^j \right) ,
\nl 
g_{W a}^{(ij)} &\equiv \frac{-g_L}{2 \sqrt{2}} ~ \left( U_{2i} ~ N_{b_1}^{j*} - V_{2i} ~ N_{b_2}^j \right),
\nl
g_{W^\prime v}^{(ij)} &\equiv \frac{g_R}{2} \left[  \frac{1}{\sqrt{2}} \left( V_{2i} ~ N_{b_1}^j + U_{2i} ~ N_{b_2}^{j*} \right) - \left( V_{1i} ~ N_t^j  + U_{1i} ~ N_t^{j*} \right) \right],
\nl
g_{W^\prime a}^{(ij)} &\equiv \frac{-g_R}{2} \left[  \frac{1}{\sqrt{2}} \left( V_{2i} ~ N_{b_1}^j - U_{2i} ~ N_{b_2}^{j*} \right) - \left( V_{1i} ~ N_t^j  - U_{1i} ~ N_t^{j*} \right) \right],
\nl
g_{Zv}^{(i +)} &\equiv  \frac{-g_L}{4 c_w} \left( U_{1i}^2 + V_{1i}^2 - 2 ~ c_{2 w} \right) ,
\nl
g_{Za}^{(i +)} &\equiv  \frac{-g_L}{4 c_w} \left( U_{1i}^2 - V_{1i}^2\right),
\nl
g_{Z^\prime v}^{(i +)} &\equiv  \frac{1}{4} \sqrt{g_R^2 - g^{\prime 2}} \left(  U_{1i}^2 + V_{1i}^2 + 2 \right),
\end{align}

\begin{align}
g_{Z^\prime a}^{(i +)} &\equiv  \frac{1}{4} \sqrt{g_R^2 - g^{\prime 2}} \left(  U_{1i}^2 - V_{1i}^2\right),
\nl
 g_{Zv}^{(i j +)}  &\equiv \frac{-g_L}{4 c_w} \left( U_{1i} ~ U_{1j} + V_{1i} ~ V_{1j} \right),
 \nl  
 g_{Za}^{(i j +)}  &\equiv \frac{-g_L}{4 c_w} \left( U_{1i} ~ U_{1j} - V_{1i} ~ V_{1j} \right),
 \nl
 g_{Z^\prime v}^{(i j +)}  &\equiv \frac{1}{4} \sqrt{g_R^2 - g^{\prime 2}} ~ \left( U_{1i} ~ U_{1j} + V_{1i} ~ V_{1j} \right),
 \nl  
 g_{Z^\prime a}^{(i j +)}  &\equiv \frac{1}{4} \sqrt{g_R^2 - g^{\prime 2}} ~ \left( U_{1i} ~ U_{1j} - V_{1i} ~ V_{1j} \right).
\end{align}

Similarly, the dark sector possesses the following interactions with the Higgs bosons of the bidoublet:
\begin{align}
\mathcal{L}_{\x \phi} &=  \sum\limits_i ~ \left( \lambda_h^{(i)}  ~ h + \lambda_H^{(i)}  ~ H \right) ~ \xiibar \xii + \lambda_A^{(i)}  ~ A ~ \xiibar i \gamma^5 \xii
\nl
&+ \sum\limits_i ~ \left( \lambda_h^{(+i)}  ~ h + \lambda_H^{(+i)}  ~ H \right) ~ \overline{\xp_i} \xp_i + \lambda_A^{(+i)}  ~ A ~ \overline{\xp_i} i \gamma^5 \xp_i
\nl
&+ \sum\limits_{i < j} ~ h ~ \xiibar \left(  \lambda_{hs}^{(ij)} + \lambda_{hp}^{(ij)} ~ i \gamma^5 \right) \xjj + H ~ \xiibar \left(  \lambda_{Hs}^{(ij)} + \lambda_{Hp}^{(ij)} ~ i \gamma^5 \right) \xjj +A ~ \xiibar \left(  \lambda_{As}^{(ij)} + \lambda_{Ap}^{(ij)} ~ i \gamma^5 \right) \xjj 
\nl
&+\big\{  \sum\limits_{i < j} ~ h ~ \overline{\xp_i} \left(  \lambda_{hs}^{(+ij)} + \lambda_{hp}^{(+ij)} ~ \gamma^5 \right) \xp_j + H ~ \overline{\xp_i} \left(  \lambda_{Hs}^{(+ij)} + \lambda_{Hp}^{(+ij)} ~ \gamma^5 \right) \xp_j 
\nl
&~~~~~~~~~~~+ i A ~ \overline{\xp_i} \left(  \lambda_{As}^{(+ij)} + \lambda_{Ap}^{(+ij)} ~ \gamma^5 \right) \xp_j + \text{h.c.} \big\}
\nl
&+ \sum\limits_{ij} ~ H^+ ~ \overline{\xp_i} \left( \lambda_{H^+s}^{(ij)} + \lambda_{H^+p}^{(ij)} ~ \gamma^5 \right) \xjj + H^- ~ \xjjbar \left( \lambda_{H^+s}^{(ij)*} - \lambda_{H^+p}^{(ij)*} ~ \gamma^5 \right) \xp_i,
\end{align}
where we have defined the couplings as follows:
\begin{align}
\lambda_h^{(i)} &\equiv \frac{-\lambda}{2} ~ \text{Re} \Big[  N_t^i \left( c_\beta N_{b_1}^i - s_\beta N_{b_2}^i \right) \Big] - \frac{\tilde{\lambda}}{2} ~ \text{Re} \Big[  N_t^i \left( s_\beta N_{b_1}^i - c_\beta N_{b_2}^i \right) \Big],
\nl
\lambda_H^{(i)} &\equiv \frac{-\lambda}{2} ~\text{Re} \Big[ N_t^i \left( s_\beta N_{b_1}^i + c_\beta N_{b_2}^i \right) \Big] + \frac{\tilde{\lambda}}{2} ~\text{Re} \Big[ N_t^i \left( c_\beta N_{b_1}^i + s_\beta N_{b_2}^i \right) \Big],
\nl
\lambda_A^{(i)} &\equiv \frac{-\lambda}{2} ~ \text{Re} \Big[ N_t^i \left( s_\beta N_{b_1}^i - c_\beta N_{b_2}^i \right) \Big] + \frac{\tilde{\lambda}}{2} ~ \text{Re} \Big[ N_t^i \left( c_\beta N_{b_1}^i - s_\beta N_{b_2}^i \right) \Big],
\nl
\lambda_{hs}^{(ij)} &\equiv \frac{-\lambda}{2} ~ \text{Re} \Big[ c_\beta \left( N_t^i N_{b_1}^j + N_t^j N_{b_1}^i \right) - s_\beta \left( N_t^i N_{b_2}^j + N_t^j N_{b_2}^i \right) \Big]
\nl
& - \frac{\tilde{\lambda}}{2} ~ \text{Re} \Big[ s_\beta \left( N_t^i N_{b_1}^j + N_t^j N_{b_1}^i \right) - c_\beta \left( N_t^i N_{b_2}^j + N_t^j N_{b_2}^i \right) \Big],
\nl
\lambda_{hp}^{(ij)} &\equiv \frac{\lambda}{2} ~ \text{Im} \Big[ c_\beta \left( N_t^i N_{b_1}^j + N_t^j N_{b_1}^i \right) - s_\beta \left( N_t^i N_{b_2}^j + N_t^j N_{b_2}^i \right) \Big]
\nl
& + \frac{\tilde{\lambda}}{2} ~ \text{Im} \Big[ s_\beta \left( N_t^i N_{b_1}^j + N_t^j N_{b_1}^i \right) - c_\beta \left( N_t^i N_{b_2}^j + N_t^j N_{b_2}^i \right) \Big],
\end{align}

\begin{align}
\lambda_{Hs}^{(ij)} &\equiv \frac{-\lambda}{2} ~ \text{Re} \Big[ s_\beta \left( N_t^i N_{b_1}^j + N_t^j N_{b_1}^i \right) + c_\beta \left( N_t^i N_{b_2}^j + N_t^j N_{b_2}^i \right) \Big] 
\nl
& + \frac{\tilde{\lambda}}{2} ~ \text{Re} \Big[ c_\beta \left( N_t^i N_{b_1}^j + N_t^j N_{b_1}^i \right) + s_\beta \left( N_t^i N_{b_2}^j + N_t^j N_{b_2}^i \right) \Big],
\nl
\lambda_{Hp}^{(ij)} &\equiv \frac{\lambda}{2} ~ \text{Im} \Big[ s_\beta \left( N_t^i N_{b_1}^j + N_t^j N_{b_1}^i \right) + c_\beta \left( N_t^i N_{b_2}^j + N_t^j N_{b_2}^i \right) \Big] 
\nl
& - \frac{\tilde{\lambda}}{2} ~ \text{Im} \Big[ c_\beta \left( N_t^i N_{b_1}^j + N_t^j N_{b_1}^i \right) + s_\beta \left( N_t^i N_{b_2}^j + N_t^j N_{b_2}^i \right) \Big],
\nl
\lambda_{As}^{(ij)} &\equiv \frac{-\lambda}{2} ~ \text{Im} \Big[ s_\beta \left( N_t^i N_{b_1}^j + N_t^j N_{b_1}^i \right) - c_\beta \left( N_t^i N_{b_2}^j + N_t^j N_{b_2}^i \right) \Big] 
\nl
& + \frac{\tilde{\lambda}}{2} ~ \text{Im} \Big[ c_\beta \left( N_t^i N_{b_1}^j + N_t^j N_{b_1}^i \right) - s_\beta \left( N_t^i N_{b_2}^j + N_t^j N_{b_2}^i \right) \Big],
\nl
\lambda_{Ap}^{(ij)} &\equiv \frac{-\lambda}{2} ~ \text{Re} \Big[ s_\beta \left( N_t^i N_{b_1}^j + N_t^j N_{b_1}^i \right) - c_\beta \left( N_t^i N_{b_2}^j + N_t^j N_{b_2}^i \right) \Big] 
\nl
& + \frac{\tilde{\lambda}}{2} ~ \text{Re} \Big[ c_\beta \left( N_t^i N_{b_1}^j + N_t^j N_{b_1}^i \right) - s_\beta \left( N_t^i N_{b_2}^j + N_t^j N_{b_2}^i \right) \Big],
\nl
\lambda_h^{(+i)} &\equiv \frac{\lambda}{\sqrt{2}} ~ \left( c_\beta ~ U_{1i} V_{2i} - s_\beta ~ U_{2i} V_{1i} \right) + \frac{\tilde{\lambda}}{\sqrt{2}} ~ \left( s_\beta ~ U_{1i} V_{2i} - c_\beta ~ U_{2i} V_{1i} \right),
\nl
\lambda_H^{(+i)} &\equiv \frac{\lambda}{\sqrt{2}} ~ \left( s_\beta ~ U_{1i} V_{2i} + c_\beta ~ U_{2i} V_{1i} \right) - \frac{\tilde{\lambda}}{\sqrt{2}} ~ \left( c_\beta ~ U_{1i} V_{2i} + s_\beta ~ U_{2i} V_{1i} \right),
\nl
\lambda_A^{(+i)} &\equiv \frac{\lambda}{\sqrt{2}} ~ \left( s_\beta ~ U_{1i} V_{2i} - c_\beta ~ U_{2i} V_{1i} \right) - \frac{\tilde{\lambda}}{\sqrt{2}} ~ \left( c_\beta ~ U_{1i} V_{2i} - s_\beta ~ U_{2i} V_{1i} \right),
\nl
\lambda_G^{(+i)} &\equiv \frac{-\lambda}{\sqrt{2}} ~ \left( c_\beta ~ U_{1i} V_{2i} + s_\beta ~ U_{2i} V_{1i} \right) - \frac{\tilde{\lambda}}{\sqrt{2}} ~ \left( s_\beta ~ U_{1i} V_{2i} + c_\beta ~ U_{2i} V_{1i} \right),
\nl
\lambda_{hs}^{(+ij)} &\equiv \frac{\lambda}{2 \sqrt{2}} ~ \Big[ c_\beta \left( U_{1i} V_{2j} + U_{1j} V_{2i} \right) - s_\beta \left( U_{2i} V_{1j} + U_{2j} V_{1i} \right) \Big]
\nl
&+ \frac{\tilde{\lambda}}{2 \sqrt{2}} ~ \Big[ s_\beta \left( U_{1i} V_{2j} + U_{1j} V_{2i} \right) - c_\beta \left( U_{2i} V_{1j} + U_{2j} V_{1i} \right) \Big],
\nl
\lambda_{hp}^{(+ij)} &\equiv \frac{\lambda}{2 \sqrt{2}} ~ \Big[ - c_\beta \left( U_{1i} V_{2j} - U_{1j} V_{2i} \right) + s_\beta \left( U_{2i} V_{1j} - U_{2j} V_{1i} \right) \Big]
\nl
&+ \frac{\tilde{\lambda}}{2 \sqrt{2}} ~ \Big[ - s_\beta \left( U_{1i} V_{2j} - U_{1j} V_{2i} \right) + c_\beta \left( U_{2i} V_{1j} - U_{2j} V_{1i} \right) \Big],
\nl
\lambda_{Hs}^{(+ij)} &\equiv \frac{\lambda}{2 \sqrt{2}} ~ \Big[ s_\beta \left( U_{1i} V_{2j} + U_{1j} V_{2i} \right) + c_\beta \left( U_{2i} V_{1j} + U_{2j} V_{1i} \right) \Big]
\nl
&- \frac{\tilde{\lambda}}{2 \sqrt{2}} ~ \Big[ c_\beta \left( U_{1i} V_{2j} + U_{1j} V_{2i} \right) + s_\beta \left( U_{2i} V_{1j} + U_{2j} V_{1i} \right) \Big],
\nl
\lambda_{Hp}^{(+ij)} &\equiv \frac{-\lambda}{2 \sqrt{2}} ~ \Big[ s_\beta \left( U_{1i} V_{2j} - U_{1j} V_{2i} \right) + c_\beta \left( U_{2i} V_{1j} - U_{2j} V_{1i} \right) \Big]
\nl
&+ \frac{\tilde{\lambda}}{2 \sqrt{2}} ~ \Big[ c_\beta \left( U_{1i} V_{2j} - U_{1j} V_{2i} \right) + s_\beta \left( U_{2i} V_{1j} - U_{2j} V_{1i} \right) \Big],
\end{align}
\begin{align}
\lambda_{As}^{(+ij)} &\equiv \frac{\lambda}{2 \sqrt{2}} ~ \Big[ - s_\beta \left( U_{1i} V_{2j} - U_{1j} V_{2i} \right) + c_\beta \left( U_{2i} V_{1j} - U_{2j} V_{1i} \right) \Big]
\nl
&+ \frac{\tilde{\lambda}}{2 \sqrt{2}} ~ \Big[ c_\beta \left( U_{1i} V_{2j} - U_{1j} V_{2i} \right) - s_\beta \left( U_{2i} V_{1j} - U_{2j} V_{1i} \right) \Big],
\nl
\lambda_{Ap}^{(+ij)} &\equiv \frac{\lambda}{2 \sqrt{2}} ~ \Big[ s_\beta \left( U_{1i} V_{2j} + U_{1j} V_{2i} \right) - c_\beta \left( U_{2i} V_{1j} + U_{2j} V_{1i} \right) \Big]
\nl
&+ \frac{\tilde{\lambda}}{2 \sqrt{2}} ~ \Big[ - c_\beta \left( U_{1i} V_{2j} + U_{1j} V_{2i} \right) + s_\beta \left( U_{2i} V_{1j} + U_{2j} V_{1i} \right) \Big],
\nl
\lambda_{H^+s}^{(ij)} &\equiv \frac{-\lambda}{2} \Big[ c_\beta \big( V_{1i} N_{b_1}^{j*} + \frac{1}{\sqrt{2}} ~ V_{2i} N_t^{j*} \big) + s_\beta \big( U_{1i} N_{b_2}^{j} + \frac{1}{\sqrt{2}} ~ U_{2i} N_t^{j} \big) \Big]
\nl
&+ \frac{\tilde{\lambda}}{2} \Big[ s_\beta \big( V_{1i} N_{b_1}^{j*} + \frac{1}{\sqrt{2}} ~ V_{2i} N_t^{j*} \big) + c_\beta \big( U_{1i} N_{b_2}^{j} + \frac{1}{\sqrt{2}} ~ U_{2i} N_t^{j} \big) \Big]
\nl
\lambda_{H^+p}^{(ij)} &\equiv \frac{\lambda}{2} \Big[ - c_\beta \big( V_{1i} N_{b_1}^{j*} + \frac{1}{\sqrt{2}} ~ V_{2i} N_t^{j*} \big) + s_\beta \big( U_{1i} N_{b_2}^{j} + \frac{1}{\sqrt{2}} ~ U_{2i} N_t^{j} \big) \Big],
\nl
&+ \frac{\tilde{\lambda}}{2} \Big[ s_\beta \big( V_{1i} N_{b_1}^{j*} + \frac{1}{\sqrt{2}} ~ V_{2i} N_t^{j*} \big) - c_\beta \big( U_{1i} N_{b_2}^{j} + \frac{1}{\sqrt{2}} ~ U_{2i} N_t^{j} \big) \Big].
\end{align}
%

\section{Loop Corrections to Mass Splittings}
\label{sec:loops}

In calculating the radiative corrections to the mass splitting, $m_{\x_1^\pm} - m_{\x_1}$, we work in the tree-level mass eigenstate basis, utilizing the $\overline{\text{MS}}$ scheme in Feynman gauge. Following the discussion in Sec.~6 of Ref.~\cite{Hirsch:2000ef}, the loop-corrected mass matrix for the neutral or charged states is given by:
\begin{align}
M_{ij} &\approx m_i ~ \delta_{ij} - \frac{1}{2} \left[ \Sigma^s_{ij} (m_i^2) + \Sigma^s_{ij} (m_j^2) + m_i ~ \Sigma^p_{ij} (m_i^2) + m_j ~ \Sigma^p_{ij} (m_j^2) \right]
\nl
\Sigma &= \Sigma^s + \slashed{p} ~ \Sigma^p + \cdots
~,
\end{align}
where the ellipsis denotes $\gamma^5$ terms, $i \Sigma (p)$ is the amputated self-energy loop diagram, $p$ is the external momentum, and $m_i$ are the tree-level mass eigenvalues. $\Sigma_{ij}$ denotes an outgoing (incoming) field $\x_i$ ($\x_j$). Similarly, $\Sigma_{ij}^\pm$ corresponds to an outgoing (incoming) field $\x_i^\pm$ ($\x_j^\pm$). The loop-corrected masses are then found by diagonalizing $M$, or in the case of charged Dirac fermions, $M^\dagger M\, $. For later convenience, we will define the generalized Feynman integral,
\begin{align}
I^{(n)} (p^2, m_1, m_2, \mu) &= \text{Re} \int_0^1 dx ~ x^n ~ \log{\left(\frac{\mu^2}{x(x-1)p^2 + (1-x) m_1^2+x \, m_2^2}\right)}
~,
\end{align}
where $\mu$ is the $\overline{\text{MS}}$ renormalization scale. In our numerical analysis, we set $\mu = 1$ TeV throughout.

In each of the contributions below, we parametrize the Lagrangian governing an individual loop diagram in a generic manner. Hence, these results can easily be mapped onto the specific models of this paper by comparing to the particular couplings in Appendices~\ref{appA}-\ref{appD}. Furthermore, since our forms are sufficiently general, they may be used to calculate radiative mass corrections in most models that incorporate vector and scalar interactions with Majorana and Dirac fermions.

\subsection{Diagonal Self-Energies for Majorana Fermions}

In this subsection, we present individual loop contributions to the $\x_i-\x_i$ self-energy, $i \Sigma_{ii}(p)$, where $\x_i$ is any Majorana fermion of a given model. Fields that are solely present in the loop are given different names depending on the nature of the interaction, e.g., $\x_j$, $\x^\pm$, $\phi$, $V$, etc. Below, the notation $\delta_{\x_i \x_j}$ denotes a Kronecker delta that is equal to unity if the Majorana fermion $\x_i$ is identical to $\x_j$ and zero otherwise.

For a loop consisting of a neutral fermion, $\x_j$, and a neutral scalar, $\phi$, we parametrize the Lagrangian as:
\be
\mathcal{L} \supset \phi ~ \bar{\x}_i (\lambda_s + \lambda_p i \gamma^5) \x_j
~.
\ee
The contribution to $\Sigma_{ii} (p)$ is given by:
\be
\Sigma_{ii} (p) = \frac{1 + 3 \delta_{\x_i \x_j}}{16 \pi ^2} \left[ \lambda^{(-)} ~ m_{\x_j}  ~ I^{(0)} (p^2, m_{\x_j}, m_\phi, \mu) + \lambda^{(+)} \slashed{p} ~ I^{(1)} (p^2, m_{\x_j}, m_\phi, \mu)  \right]
~,
\ee
where 
\be
\lambda^{(\pm)} \equiv \lambda_s^2 \pm \lambda_p^2
~.
\ee
The analogous charged loop consisting of a charged fermion, $\x^\pm$, and charged scalar, $\phi^\pm$, that interact through
\be
\mathcal{L} \supset \phi^+ ~ \xpbar (\lambda_s + \lambda_p \gamma^5) \x_i + \text{h.c.}
~,
\ee
is given by:
\be
\Sigma_{ii} (p) = \frac{1}{8 \pi ^2} \left[ \lambda^{(-)} ~ m_{\x^\pm} ~ I^{(0)}(p^2, m_{\x^\pm}, m_{\phi^\pm}, \mu) + \lambda^{(+)} ~ \slashed{p} ~ I^{(1)}(p^2, m_{\x^\pm}, m_{\phi^\pm}, \mu) \right]
~,
\ee
where 
\be
\lambda^{(\pm)} \equiv |\lambda_s|^2 \pm |\lambda_p|^2
~.
\ee

When a neutral fermion, $\x_j$, and a neutral vector, $V^\mu$, run in the loop, we parametrize the Lagrangian as:
\be
\mathcal{L} \supset V_\mu ~ \bar{\x}_i  \gamma^\mu ( i g_v + g_a \gamma^5) \x_j
~.
\ee
This contributes to the self-energy as:
\begin{align}
\Sigma_{ii} (p) &= \frac{1+3 \delta_{\x_i \x_j}}{16 \pi ^2} \Bigg\{  2 g^{(-)} ~ m_{\x_j} ~ \left[ 1 - 2 I^{(0)}(p^2, m_{\x_j}, m_V, \mu) \right] 
\nl
& \qquad \qquad \quad ~~~
+ g^{(+)} ~ \slashed{p} ~ \left[ 2 I^{(1)}(p^2, m_{\x_j}, m_V, \mu) - 1  \right] \Bigg\}
~,
\end{align}
where 
\be
g^{(\pm)} \equiv g_v^2 \pm g_a^2
~.
\ee
Similarly, in the case of a charged loop made up of a charged fermion, $\x^\pm$, and a charged vector, $V_\mu^\pm$, the Lagrangian is given by:
\be
\mathcal{L} \supset V_\mu^+ ~ \overline{\x^+}  \gamma^\mu ( g_v + g_a \gamma^5) \x_i + \text{h.c.}
~,
\ee
and the self-energy contribution is found to be:
\begin{align}
\Sigma_{ii} (p) &= \frac{1}{8 \pi ^2} \Bigg\{  2 g^{(-)} ~ m_{\x^\pm} ~ \left[ 1 - 2 I^{(0)}(p^2, m_{\x^\pm}, m_{V^\pm}, \mu) \right] 
\nl
& \qquad \quad 
+ g^{(+)} ~ \slashed{p} ~ \left[ 2 I^{(1)}(p^2, m_{\x^\pm}, m_{V^\pm}, \mu) - 1  \right] \Bigg\}
~,
\end{align}
where 
\be
g^{(\pm)} \equiv |g_v|^2 \pm |g_a|^2
~.
\ee
%

\subsection{Off-Diagonal Self-Energies for Majorana Fermions}

In this subsection, we present individual loop contributions to the off-diagonal $\x_i-\x_j$ self-energy ($i \neq j$), $i \Sigma_{ji}(p)$, where $\x_i$ and $\x_j$ are any distinct Majorana fermions of a given model. Fields that are solely present in the loop are given different names depending on the nature of the interaction, e.g., $\x_k$, $\x^\pm$, $\phi$, $V$, etc. Below, the notation $\delta_{\x_i \x_j}$ denotes a Kronecker delta that is equal to unity if the Majorana fermion $\x_i$ is identical to $\x_j$ and zero otherwise. Furthermore, we use the notation where $\lambda_{\{ 1} \tilde{\lambda}_{2 \}} \equiv \lambda_1 \tilde{\lambda}_2 + \lambda_2 \tilde{\lambda}_1$ and $\lambda_{[ 1} \tilde{\lambda}_{2 ]} \equiv \lambda_1 \tilde{\lambda}_2 - \lambda_2 \tilde{\lambda}_1$.

For a loop consisting of a neutral fermion, $\x_k$, and a neutral scalar, $\phi$, the Lagrangian is given by:
\be
\mathcal{L} \supset \phi ~ \bar{\x}_i (\lambda_s^{(ik)} + \lambda_p^{(ik)} i \gamma^5) \x_k + \phi ~ \bar{\x}_k (\lambda_s^{(kj)} + \lambda_p^{(kj)} i \gamma^5) \x_j
~,
\ee
and the contribution to the self-energy is
\begin{align}
\Sigma_{ji} (p) &= \frac{( 1 + \delta_{\x_i \x_k})( 1 + \delta_{\x_j \x_k})}{16 \pi ^2} \bigg[ m_{\x_k} ~ \left( \lambda^{(-)} + \lambda^{(ik)}_{\{ s} \lambda^{(kj)}_{p \}} ~ i \gamma^5 \right)  ~ I^{(0)} (p^2, m_{\x_k}, m_\phi, \mu) 
\nl
&~~~~~~~~~~~~~~~~~~~~~~~~~~~~~~~~~~
+ \slashed{p} ~ \left( \lambda^{(+)} - \lambda^{(ik)}_{[ s} \lambda^{(kj)}_{p ]} ~ i \gamma^5 \right)  ~ I^{(1)} (p^2, m_{\x_k}, m_\phi, \mu)  \bigg]
~,
\end{align}
where 
\be
\lambda^{(\pm)} \equiv \lambda_s^{(ik)} \lambda_s^{(kj)} \pm \lambda_p^{(ik)} \lambda_p^{(kj)}
~.
\ee
Similarly, there is a charged loop consisting of a charged fermion, $\x^\pm$, and charged scalar, $\phi^\pm$. We parametrize the Lagrangian as:
\be
\mathcal{L} \supset \phi^+ ~ \xpbar (\lambda_s^{(i)} + \lambda_p^{(i)} \gamma^5) \x_i + \phi^+ ~ \xpbar (\lambda_s^{(j)} + \lambda_p^{(j)} \gamma^5) \x_j + \text{h.c.}
~.
\ee
The self-energy is then given by:
\begin{align}
\Sigma_{ji} (p) &= \frac{1}{8 \pi ^2} \bigg[ m_{\x^\pm} ~ \left( \text{Re} \lambda^{(-)} - \text{Im} \lambda^{(i)}_{[ s} \lambda^{(j)*}_{p ]} ~ i \gamma^5 \right)  ~ I^{(0)} (p^2, m_{\x^\pm}, m_{\phi^\pm}, \mu) 
\nl
&~~~~~~~~~~~~
+ \slashed{p} ~ \left( \text{Re} \lambda^{(+)} + \text{Im} \lambda^{(i)}_{\{ s} \lambda^{(j)*}_{p \}} ~ i \gamma^5 \right)  ~ I^{(1)} (p^2, m_{\x^\pm}, m_{\phi^\pm}, \mu)  \bigg]
~,
\end{align}
where 
\be
\lambda^{(\pm)} \equiv \lambda_s^{(i)} \lambda_s^{(j)*} \pm \lambda_p^{(i)} \lambda_p^{(j)*}
~.
\ee

When a neutral fermion, $\x_k$, and a neutral vector, $V^\mu$, run in the loop, the Lagrangian is given by:
\be
\mathcal{L} \supset V_\mu ~ \bar{\x}_i  \gamma^\mu ( i g_v^{(ik)} + g_a^{(ik)} \gamma^5) \x_k + V_\mu ~ \bar{\x}_k  \gamma^\mu ( i g_v^{(kj)} + g_a^{(kj)} \gamma^5) \x_j
~.
\ee
For the self-energy, we find:
\begin{align}
\Sigma_{ji} (p) &= \frac{-( 1 + \delta_{\x_i \x_k})( 1 + \delta_{\x_j \x_k})}{16 \pi ^2} \bigg[ 2 m_{\x_k} ~ \left( g^{(+)} - g^{(ik)}_{[ v} g^{(kj)}_{a ]} ~ i  \gamma^5 \right)  ~ \left( 2 I^{(0)} (p^2, m_{\x_k}, m_V, \mu) - 1 \right)
\nl
&~~~~~~~~~~~~~~~~~~~~~~
+ \slashed{p} ~ \left( g^{(-)} + g^{(ik)}_{\{ v} g^{(kj)}_{a \}} ~ i  \gamma^5 \right)  ~ \left( 1 - 2 I^{(1)} (p^2, m_{\x_k}, m_V, \mu) \right)  \bigg]
~,
\end{align}
where 
\be
g^{(\pm)} \equiv g_v^{(ik)} g_v^{(kj)} \pm g_a^{(ik)} g_a^{(kj)}
~.
\ee
There is also a similar contribution from an intermediate charged fermion, $\x^\pm$, and a charged vector, $V_\mu^\pm$. We parametrize the Lagrangian as:
\be
\mathcal{L} \supset V_\mu^+ ~ \overline{\x^+}  \gamma^\mu ( g_v^{(i)} + g_a^{(i)} \gamma^5) \x_i + V_\mu^+ ~ \overline{\x^+}  \gamma^\mu ( g_v^{(j)} + g_a^{(j)} \gamma^5) \x_j  + \text{h.c.}
~.
\ee
We then find for the self-energy:
\begin{align}
\Sigma_{ji} (p) &= \frac{1}{8 \pi ^2} \bigg[ 2 m_{\x^\pm} ~ \left( \text{Re} g^{(-)} - \text{Im} g^{(i)}_{[ v} g^{(j)*}_{a ]} ~ i \gamma^5 \right)  ~ \left( 1 - 2 I^{(0)} (p^2, m_{\x^\pm}, m_{V^\pm}, \mu) \right)
\nl
&~~~~~~~~~~~~~
+ \slashed{p} ~ \left( \text{Re} g^{(+)} + \text{Im} g^{(i)}_{\{ v} g^{(j)*}_{a \}} ~ i \gamma^5 \right)  ~ \left( 2 I^{(1)} (p^2, m_{\x^\pm}, m_{V^\pm}, \mu) - 1 \right)  \bigg]
~,
\end{align}
where 
\be
g^{(\pm)} \equiv g_v^{(i)} g_v^{(j)*} \pm g_a^{(i)} g_a^{(j)*}
~.
\ee
%

\subsection{Diagonal Self-Energies for Dirac Fermions}

In this subsection, we present individual loop contributions to the $\x_i^\pm-\x_i^\pm$ self-energy, $i \Sigma_{ii}^\pm(p)$, where $\x_i^\pm$ is any charged Dirac fermion of a given model. Fields that are solely present in the loop are given different names depending on the nature of the interaction, e.g., $\x$, $\x_j^\pm$, $\phi$, $V$, etc.

For the loop contribution consisting of a charged fermion, $\x_j^\pm$, and a neutral scalar, $\phi$, the Lagrangian is given by:
\be
\mathcal{L} \supset \phi ~ \overline{\x_i^+} (\lambda_s + \lambda_p \gamma^5) \x_j^+ + \text{h.c.}
~,
\ee
and the self-energy is found to be:
\begin{align}
\Sigma_{ii}^\pm (p) &= \frac{1}{16 \pi ^2} \Big[ \lambda^{(-)} ~ m_{\x_j^\pm}  ~ I^{(0)} (p^2, m_{\x_j^\pm}, m_\phi, \mu) 
\nl
&\qquad \quad
+ \slashed{p} \left( \lambda^{(+)} - 2 ~ \text{Re}\lambda_s \lambda_p^* ~ \gamma^5 \right) ~ I^{(1)} (p^2, m_{\x_j^\pm}, m_\phi, \mu)  \Big]
~,
\end{align}
where 
\be
\lambda^{(\pm)} \equiv |\lambda_s|^2 \pm |\lambda_p|^2
~.
\ee
There is a similar diagram with a neutral fermion, $\x$, and a charged scalar, $\phi^\pm$, running in the loop. The Lagrangian is given by:
\be
\mathcal{L} \supset \phi^+ ~ \overline{\x_i^+} (\lambda_s + \lambda_p \gamma^5) \x + \text{h.c.}
~,
\ee
and for the self-energy, we find:
\begin{align}
\Sigma_{ii}^\pm (p) &= \frac{1}{16 \pi ^2} \Big[ \lambda^{(-)} ~ m_\x  ~ I^{(0)} (p^2, m_\x, m_\phi^\pm, \mu) 
\nl
& \qquad \quad
+ \slashed{p} \left( \lambda^{(+)} - 2 ~ \text{Re} \lambda_s \lambda_p^* ~ \gamma^5 \right) ~ I^{(1)} (p^2, m_\x, m_\phi^\pm, \mu)  \Big]
~,
\end{align}
where 
\be
\lambda^{(\pm)} \equiv |\lambda_s|^2 \pm |\lambda_p|^2
~.
\ee

We now consider the contribution from a loop consisting of a charged fermion, $\x_j^\pm$, and a neutral vector, $V^\mu$. The Lagrangian is given by:
\be
\mathcal{L} \supset V_\mu ~ \overline{\x_i^+}  \gamma^\mu ( g_v + g_a \gamma^5) \x_j^+ + \text{h.c.}
~,
\ee
and the self-energy is
\begin{align}
\Sigma_{ii}^\pm (p) &= \frac{1}{16 \pi ^2} \Bigg\{  2 g^{(-)} ~ m_{\x_j^\pm} ~ \left[ 1 - 2 I^{(0)}(p^2, m_{\x_j^\pm}, m_V, \mu) \right] 
\nl
&~~~~~~~~~~~~ + \slashed{p} \left( g^{(+)} + 2 ~ g_v g_a ~ \gamma^5 \right) ~ \left[ 2 I^{(1)}(p^2, m_{\x_j^\pm}, m_V, \mu) - 1  \right] \Bigg\}
~,
\end{align}
where 
\be
g^{(\pm)} \equiv g_v^2 \pm g_a^2
~.
\ee
In the case that a neutral fermion, $\x$, and a charged vector, $V_\mu^\pm$, run in the loop, we parametrize the Lagrangian as:
\be
\mathcal{L} \supset V_\mu^+ ~ \overline{\x_i^+}  \gamma^\mu ( g_v + g_a \gamma^5) \x + \text{h.c.}
~,
\ee
which gives:
\begin{align}
\Sigma_{ii}^\pm (p) &= \frac{1}{16 \pi ^2} \Bigg\{  2 g^{(-)} ~ m_\x ~ \left[ 1 - 2 I^{(0)}(p^2, m_{\x}, m_{V^\pm}, \mu) \right] 
\nl
&~~~~~~~~~~~~
+\slashed{p} \left( g^{(+)} + 2 ~ \text{Re} g_v g_a^* ~ \gamma^5 \right) ~ \left[ 2 I^{(1)}(p^2, m_\x, m_{V^\pm}, \mu) - 1  \right] \Bigg\}
~,
\end{align}
where 
\be
g^{(\pm)} \equiv |g_v|^2 \pm |g_a|^2
~.
\ee
%

\subsection{Off-Diagonal Self-Energies for Dirac Fermions}

Here, we present individual loop contributions to the off-diagonal $\x_i^\pm-\x_j^\pm$ self-energy ($i \neq j$), $i \Sigma_{ji}^\pm(p)$, where $\x_i^\pm$ and $\x_j^\pm$ are any distinct charged Dirac fermions of a given model. Fields that are solely present in the loop are given different names depending on the nature of the interaction, e.g., $\x_k^\pm$, $\x$, $\phi$, $V$, etc. We use the notation where $\lambda_{\{ 1} \tilde{\lambda}_{2 \}} \equiv \lambda_1 \tilde{\lambda}_2 + \lambda_2 \tilde{\lambda}_1$ and $\lambda_{[ 1} \tilde{\lambda}_{2 ]} \equiv \lambda_1 \tilde{\lambda}_2 - \lambda_2 \tilde{\lambda}_1$.

We first consider the loop consisting of a charged fermion, $\x_k^\pm$, and a neutral scalar, $\phi$. The couplings are parametrized as:
\be
\mathcal{L} \supset \phi ~ \overline{\x_i^+} (\lambda_s^{(ik)} + \lambda_p^{(ik)} \gamma^5) \x_k^+ + \phi ~ \overline{\x_k^+} (\lambda_s^{(kj)} + \lambda_p^{(kj)} \gamma^5) \x_j^+ + \text{h.c.}
~.
\ee
We then find for the self-energy:
\begin{align}
\Sigma_{ji}^\pm (p) &= \frac{1}{16 \pi ^2} \bigg[ m_{\x_k^\pm} ~ \left( \lambda^{(+)*} - \lambda^{(ik)*}_{\{ s} \lambda^{(kj)*}_{p \}} ~ \gamma^5 \right)  ~ I^{(0)} (p^2, m_{\x_k^\pm}, m_\phi, \mu) 
\nl
&~~~~~~~~~~~~~
+ \slashed{p} ~ \left( \lambda^{(-)*} + \lambda^{(ik)*}_{[ s} \lambda^{(kj)*}_{p ]} ~ \gamma^5 \right)  ~ I^{(1)} (p^2, m_{\x_k^\pm}, m_\phi, \mu)  \bigg]
~,
\end{align}
where 
\be
\lambda^{(\pm)} \equiv \lambda_s^{(ik)} \lambda_s^{(kj)} \pm \lambda_p^{(ik)} \lambda_p^{(kj)}
~.
\ee
For the similar diagram with a neutral fermion, $\x$, and a charged scalar, $\phi^\pm$, the Lagrangian is given by:
\be
\mathcal{L} \supset \phi^+ ~ \overline{\x_i^+} (\lambda_s^{(i)} + \lambda_p^{(i)} \gamma^5) \x + \phi^+ ~ \overline{\x_j^+} (\lambda_s^{(j)} + \lambda_p^{(j)} \gamma^5) \x + \text{h.c.}
~.
\ee
For the self-energy, we then find:
\begin{align}
\Sigma_{ji}^\pm (p) &= \frac{1}{16 \pi ^2} \bigg[ m_\x ~ \left( \lambda^{(-)*} + \lambda^{(i)*}_{[ s} \lambda^{(j)}_{p ]} ~ \gamma^5 \right)  ~ I^{(0)} (p^2, m_{\x}, m_{\phi^\pm}, \mu) 
\nl
&~~~~~~~~~~~~
+ \slashed{p} ~ \left( \lambda^{(+)*} - \lambda^{(i)*}_{\{ s} \lambda^{(j)}_{p \}} ~ \gamma^5 \right)  ~ I^{(1)} (p^2, m_{\x}, m_{\phi^\pm}, \mu)  \bigg]
~,
\end{align}
where 
\be
\lambda^{(\pm)} \equiv \lambda_s^{(i)} \lambda_s^{(j)*} \pm \lambda_p^{(i)} \lambda_p^{(j)*}
~.
\ee

When a charged fermion, $\x_k^\pm$, and a neutral vector, $V^\mu$, run in the loop, we parametrize the Lagrangian as:
\be
\mathcal{L} \supset V_\mu ~ \overline{\x_i^+} \gamma^\mu ( g_v^{(ik)} + g_a^{(ik)} \gamma^5) \x_k^+ + V_\mu ~ \overline{\x_k^+}  \gamma^\mu ( g_v^{(kj)} + g_a^{(kj)} \gamma^5) \x_j^+ + \text{h.c.}
~.
\ee
The contribution to the self-energy is given by:
\begin{align}
\Sigma_{ji}^\pm (p) &= \frac{1}{16 \pi ^2} \bigg[ 2 m_{\x_k^\pm} ~ \left( g^{(-)*} - g^{(ik)*}_{[ v} g^{(kj)*}_{a ]} ~  \gamma^5 \right)  ~ \left( 1 - 2 I^{(0)} (p^2, m_{\x_k^\pm}, m_V, \mu) \right)
\nl
&~~~~~~~~~~~~~~~~~~~~~~~~~~~~~~~~~~
+ \slashed{p} ~ \left( g^{(+)*} + g^{(ik)*}_{\{ v} g^{(kj)*}_{a \}} ~ \gamma^5 \right)  ~ \left( 2 I^{(1)} (p^2, m_{\x_k^\pm}, m_V, \mu) - 1 \right)  \bigg]
~,
\end{align}
where 
\be
g^{(\pm)} \equiv g_v^{(ik)} g_v^{(kj)} \pm g_a^{(ik)} g_a^{(kj)}
~.
\ee
There is also a diagram with a neutral fermion, $\x$, and a charged vector, $V_\mu^\pm$, in the loop. The Lagrangian is given by:
\be
\mathcal{L} \supset V_\mu^+ ~ \overline{\x_i^+}  \gamma^\mu ( g_v^{(i)} + g_a^{(i)} \gamma^5) \x + V_\mu^+ ~ \overline{\x_j^+}  \gamma^\mu ( g_v^{(j)} + g_a^{(j)} \gamma^5) \x + \text{h.c.}
~.
\ee
We then find:
\begin{align}
\Sigma_{ji}^\pm (p) &= \frac{1}{16 \pi ^2} \bigg[ 2 m_{\x} ~ \left( g^{(-)*} - g^{(i)*}_{[ v} g^{(j)}_{a ]} ~ \gamma^5 \right)  ~ \left( 1 - 2 I^{(0)} (p^2, m_{\x}, m_{V^\pm}, \mu) \right)
\nl
&~~~~~~~~~~~~~~~
+ \slashed{p} ~ \left( g^{(+)*} + g^{(i)*}_{\{ v} g^{(j)}_{a \}} ~ \gamma^5 \right)  ~ \left( 2 I^{(1)} (p^2, m_{\x}, m_{V^\pm}, \mu) - 1 \right)  \bigg]
~,
\end{align}
where 
\be
g^{(\pm)} \equiv g_v^{(i)} g_v^{(j)*} \pm g_a^{(i)} g_a^{(j)*}
~.
\ee

\bibliography{LRDM2.bib}
\bibliographystyle{JHEP}

\end{document}